\documentclass[twocolumn,aps,nofootinbib,longbibliography]{revtex4-1}
\usepackage[dvipdfmx]{graphicx}
\usepackage{mathtools,amssymb,amsmath,bm,url,color}
\usepackage[breaklinks]{hyperref}
\usepackage{multirow}
\allowdisplaybreaks[1]

\begin{document}

\title{
   Magnetic field-temperature phase diagrams for multiple-$Q$ magnetic orderings: \\
   Exact steepest descent approach to long-range interacting spin systems
}

\author{Yasuyuki~Kato and Yukitoshi~Motome}

\affiliation{Department of Applied Physics, the University of Tokyo, Tokyo 113-8656, Japan}

\begin{abstract}
\noindent
Multiple-$Q$ magnetic orderings represent magnetic textures composed of superpositions of multiple spin density waves or spin spirals,
as represented by two-dimensional skyrmion crystals and three-dimensional hedgehog lattices. 
Such magnetic orderings have been observed in various magnetic materials in recent years, 
and attracted enormous attention, especially from the viewpoint of topology and emergent electromagnetic fields originating from noncoplanar magnetic structures. 
Although they often exhibit successive phase transitions among different multiple-$Q$ states while changing temperature and an external magnetic field, it is not straightforward to elucidate the phase diagrams, 
mainly due to the lack of concise theoretical tools as well as appropriate microscopic models.
Here, we provide a theoretical framework for a class of effective spin models with long-range magnetic interactions mediated by conduction electrons in magnetic metals. 
Our framework is based on the steepest descent method with a set of self-consistent equations that leads to exact solutions in the thermodynamic limit, 
and has many advantages over existing methods such as biased variational calculations and numerical Monte Carlo simulations. 
We develop two methods that complement each other in terms of the computational cost and the range of applications.
As a demonstration, applying the framework to the models with instabilities toward triple- and hextuple-$Q$ magnetic orderings, 
we clarify the magnetic field-temperature phase diagrams with a variety of multiple-$Q$ phases. 
We find that the models exhibit interesting reentrant phase transitions where the multiple-$Q$ phases appear only at finite temperature and/or nonzero magnetic field. 
Furthermore, we show that the multiple-$Q$ states can be topologically-nontrivial stacked skyrmion crystals or hedgehog lattices, which exhibit large net spin scalar chirality associated with nonzero skyrmion number. 
The results demonstrate that our framework could be a versatile tool for studying magnetic and topological phase transitions and related quantum phenomena in actual magnetic metals hosting multiple-$Q$ magnetic orderings.
\end{abstract}

\maketitle

\section{Introduction}

Multiple-$Q$ magnetic orders are magnetically ordered states 
whose spin textures are approximately given by superpositions of multiple spin density waves or spin spirals.
They show mutual interference peaks in the spin structure factor in momentum space, 
which are observable in elastic neutron scattering experiments.  
In real space, they are often regarded as periodic arrays of topologically nontrivial objects made of many spins~\cite{Bogdanov1989,Braun2012,Seidel2016,Bogdanov2020,Tokura2020},
such as two-dimensional (2D) skyrmion crystals (SkXs) in triple-$Q$ ($3Q$) magnetic orderings~\cite{Muhlbauer2009,Yu2010,Nagaosa2013,Fert2017}, 
2D vortex crystals (VCs)  in double-$Q$ ($2Q$) magnetic orderings~\cite{Khanh2020},
and three-dimensional (3D) hedgehog lattices (HLs) in $3Q$ and quadruple-$Q$ magnetic orderings~\cite{Tanigaki2015,Kanazawa2016,Kanazawa2017,Fujishiro2019,Fujishiro2020,Kanazawa2020}. 
Such topological spin textures induce unique effects on electronic and transport properties through the Berry phase mechanism~\cite{Berry1984,Xiao2010}, 
such as the magnetoelectric effect~\cite{Tokura2010} and the topological Hall effect~\cite{Nagaosa2010}, 
and thus the multiple-$Q$ magnetic orderings have been attracting enormous attention for years.

Several proposals have been made for the stabilization mechanism of the multiple-$Q$ magnetic orderings, for instance, long-range dipole interactions~\cite{Lin1974,Malozemoff1979,Garel1982,Ezawa2010,Kwon2012,Utesov2021},
the Dzyaloshinskii-Moriya (DM) antisymmetric exchange interactions~\cite{Dzyaloshinsky1958,Moriya1960,Dzyaloshinskii1964,Dzyaloshinskii1965,Dzyaloshinskii1965b,Izyumov1984,Ishikawa1984,Lebech1989,Rossler2006,Kishine2009,Yi2009,Muhlbauer2009,Togawa2012,Okumura2017}, 
four-spin interactions~\cite{Momoi1997,Kurz2001,Heinze2011,Brinker2019,Laszloffy2019,Paul2020}, 
frustrated magnetic interactions~\cite{Okubo2012,Leonov2015,Lin2016}, 
and bond-dependent anisotropic interaction~\cite{Hayami2020,Hayami2021,Wang2021}.
Among them, in this paper, we focus on the long-range interactions mediated by conduction electrons~\cite{Martin2008,Akagi2010,Kato2010,Akagi2012,Ozawa2016,Ozawa2017}. 
Such interactions are incorporated into effective spin models for magnetic metals~\cite{Hayami2017,Hayami2018,Hayami2021b},
and have been shown to stabilize a variety of multiple-$Q$ magnetic orderings,
such as $2Q$ and $3Q$ VCs~\cite{Hayami2017,Hayami2018,Hayami2021c,Kato2021}
, and $3Q$ and quadruple-$Q$ HLs~\cite{Okumura2020,Shimizu2021a,Shimizu2021b,Kato2021}.
Usually, the models exhibit complicated phase diagrams while changing the lattice structures, the interaction parameters, temperature, and an external magnetic field. 
In the previous studies, such phase diagrams were studied by using, e.g., the variational method and the Monte Carlo simulation. 
It is, however, not straightforward to elucidate the phase competition between different multiple-$Q$ states. 
For instance, the variational method is basically limited to zero temperature and requires good variational states. 
The Monte Carlo simulation is an unbiased powerful tool which is applicable to not only the zero-temperature limit but also finite temperature, 
but it requires careful analysis of the finite size effect, and it is usually a time consuming task to obtain the full phase diagram because of the relatively high computational cost. 
Thus, there remain vast unexplored parameter regions, including extensions of the models to more complex multiple-$Q$ orderings, such as hextuple-$Q$ ($6Q$) ones~\cite{Binz2006a,Binz2006b,Binz2008}. 
An unbiased and computationally cheaper method is therefore highly desired.

In this paper, we develop a versatile theoretical framework for a class of the effective spin models with long-range interactions, and demonstrate its power by revealing the complete phase diagrams for two types of the models. 
Our framework is based on the self-consistent equations derived from the saddle point method, 
which gives the exact solution in the thermodynamic limit.
Specifically, we provide two methods, which we call method I and method II, being complementary to each other:
The computational cost of method I is cheaper than method II, but method II has a wider 
range of applications in terms of the interaction types.
Using this framework, we study two different models that stabilize $3Q$ and $6Q$ magnetic orderings. 
We show that both models exhibit interesting phase diagrams depending on the interaction parameters and the direction of the magnetic field. 
In particular, we find that multiple-$Q$ magnetic orderings with a larger number of components can be stabilized by raising temperature and/or applying the magnetic field, 
which yield a variety of successive and reentrant transitions in the magnetic field-temperature phase diagram.
We also show that the $3Q$ and $6Q$ states can be topologically nontrivial with nonzero spin scalar chirality.
Furthermore, we find several topological transitions associated with changes in the topological skyrmion number.

The structure of this paper is as follows.
In Sec.~\ref{sec:model}, 
we introduce the generic form of the Hamiltonian of the effective spin model for magnetic metals with long-range interactions mediated by conduction electrons. 
In Sec.~\ref{sec:method}, we describe the theoretical framework for the exact analysis of the effective spin model in the thermodynamic limit based on the steepest descent method.
We show two methods, method I and method II in Secs.~\ref{sec:method1} and \ref{sec:method2}, respectively. 
In Sec.~\ref{subsec:remark}, we give some remarks on the condition for the existence of the saddle point solutions, the computational cost, and the range of application for the two methods.
In Sec.~\ref{sec:results}, we present the results for two different models that stabilize $3Q$ and $6Q$ magnetic orderings.
For each model, after introducing the concrete model parameters (Secs.~\ref{sec:modelparam3Q} and \ref{sec:modelparam6Q}), 
we discuss the ground-state phase diagram at zero magnetic field and stable spin configurations of ground states therein (Secs.~\ref{sec:GroundState3Q} and \ref{sec:GroundState6Q}).
Then, we present the magnetic field-temperature phase diagrams for three different magnetic field directions, 
and elucidate the details of the transitions between various multiple-$Q$ phases (Secs.~\ref{sec:FiniteT3Q} and \ref{sec:FiniteT6Q}).
In Sec.~\ref{sec:hidden_transition}, we discuss two possible types of hidden transitions.
Finally, Sec.~\ref{sec:summary} is devoted to the summary and perspectives.


\section{Model}\label{sec:model}
We consider a class of spin lattice models proposed for understanding the multiple-$Q$ magnetic orderings in itinerant magnets~\cite{Hayami2017,Hayami2018,Hayami2021b}. 
The generic form of the Hamiltonian is given by spin interactions in momentum space as
\begin{align}
\mathcal{H}=\mathcal{H}({\bf S}_{{\bf Q}_1},\cdots,{\bf S}_{{\bf Q}_{N_Q}})-
  \sqrt{N} {\bf B} \cdot {\bf S}_{{\bf Q}=0} , 
\label{eq:H_general} 
\end{align}
where
\begin{align}
{\bf S}_{\bf Q} = \frac{1}{\sqrt{N}} \sum_{\bf r} {\bf S}_{\bf r} e^{- i {\bf Q} \cdot{\bf r}}.
\end{align}
Here, ${\bf S}_{\bf r}=(S^x_{\bf r},S^y_{\bf r},S^z_{\bf r})$ represents the spin degree of freedom at site ${\bf r}$ in real space, and $N$ is the total number of spins.
In this paper, we consider the classical spin limit where ${\bf S}_{\bf r} \in \mathbb{R}^3$ and $|{\bf S}_{\bf r}| = 1$, for simplicity.
In Eq.~\eqref{eq:H_general}, the first term includes the effective spin interaction mediated by itinerant electrons, where ${\bf Q}_\eta$ with $\eta=1,2,\ldots ,N_Q$ are the characteristic wave numbers given by the nesting vectors of the Fermi surfaces of itinerant electrons in the limit of weak spin-charge coupling~\cite{Hayami2017}.
Note that the interactions defined in momentum space extend over infinite distances without decay in real space. 
The second term in Eq.~\eqref{eq:H_general} represents the Zeeman coupling with an external magnetic field ${\bf B}$, described as 
$- \sum_{\bf r} {\bf B}\cdot {\bf S}_{\bf r}$ in real space.


The simplest form of the Hamiltonian in Eq.~\eqref{eq:H_general} is given by two-spin interactions in the first term as
\begin{align}
\mathcal{H} =& \sum^{N_Q}_{\eta=1} (\mathcal{H}_{{\bf Q}_\eta}+ \mathcal{H}_{-{\bf Q}_\eta}) - 
  \sqrt{N} {\bf B} \cdot {\bf S}_{{\bf Q}=0} ,
\label{eq:H}
\end{align}
where 
\begin{align}
\mathcal{H}_{{\bf Q}_\eta} =
-\sum_{\alpha,{\alpha'} } S^\alpha_{{\bf Q}_\eta} \mathcal{J}^{\alpha {\alpha'} }_{{\bf Q}_\eta} S^{\alpha'} _{-{\bf Q}_\eta}.
\label{eq:Hq2}
\end{align}
Here, $\mathcal{J}_{{\bf Q}_\eta}$ is a $3\times 3$ Hermitian matrix describing the form of the two-spin interactions;
the sums of $\alpha$ and $\alpha'$ run over $x$, $y$, and $z$. 
Note that $\mathcal{J}_{-{\bf Q}_\eta}=\mathcal{J}^*_{{\bf Q}_\eta}$ and ${\bf S}_{-{\bf Q}_\eta} = {\bf S}_{{\bf Q}_\eta}^*$.
The two-spin interactions are the lowest-order contributions in the perturbative expansion in terms of the spin-charge coupling~\cite{Hayami2017}, which include the Ruderman-Kittel-Kasuya-Yosida interaction $\propto {\bf S}_{{\bf Q}_\eta} \cdot {\bf S}_{-{\bf Q}_\eta}$~\cite{Ruderman1954,Kasuya1956,Yosida1957}.
In addition, the model can be extended to include higher-order contributions. 
For instance, 
the four-spin biquadratic interactions $\propto ({\bf S}_{{\bf Q}_\eta} \cdot {\bf S}_{-{\bf Q}_\eta})^2$~\cite{Hayami2017,Hayami2020,Hayami2020b,Okumura2020,Yasui2020,Hayami2021d,Hayami2021,Hayami2021f,Hayami2021g,Hayami2021h,Hirschberger2021,Seo2021} 
and the six-spin bicubic interactions
$\propto \{
{\bf S}_{{\bf Q}_1} \cdot (
{\bf S}_{{\bf Q}_2}
\times
{\bf S}_{{\bf Q}_3})
\}^2$~\cite{Hayami2021c}
have also been considered as the origin of stabilization of multiple-$Q$ magnetic orderings.

\section{Method}\label{sec:method}
In this section, we construct the framework to obtain the phase diagram of the model in Eq.~\eqref{eq:H_general}
based on the steepest descent method also known as the saddle-point method (see for example Ref.~\cite{Nishimori2010}).
Specifically, we develop two methods that complement each other, method I and method II.
Method I is computationally cheaper than method II, while it is only applicable to the two-spin interactions in Eq.~\eqref{eq:Hq2}. 
Meanwhile, method II has a wider range of applications; it can deal with the higher-order spin interactions.

In the following, we consider the model in Eq.~\eqref{eq:H_general} with ${\bf Q}_\eta$ being commensurate with the lattice sites. 
Since the period of magnetic ordering is set by ${\bf Q}_\eta$ as discussed in Sec.~\ref{sec:results}, 
this corresponds to considering magnetic orders commensurate with the lattice sites. 
Let ${\bf a}_\mu$ ($\mu=1$, $2,\ldots,d$) be the primitive lattice vectors in spatial dimension $d$ and ${\bf A}_\mu$ be the magnetic translation vectors spanning the magnetic unit cell:
${\bf A}_\mu = \sum_{\mu'=1}^d l_{\mu \mu'} {\bf a}_{\mu'}$ with integers $l_{\mu \mu'}$. 
Then, all the ${\bf Q}_\eta$ must satisfy
\begin{align}
e^{ i  {\bf Q}_\eta \cdot {\bf A}_\mu}=1,\label{eq:condQeta}
\end{align}
for all $\mu$.
While the magnetic unit cell can be smaller for simpler spin states such as the single-$Q$ spin state for the case of $N_Q \geq 2$, 
we take the largest common magnetic unit cell defined by ${\bf A}_\mu$.

Given such commensurate situations, it is commonly useful to denote the spatial coordinate ${\bf r}$ as ${\bf r}={\bf R} + {\bf r}_0$
where ${\bf R}$ and ${\bf r}_0$ are the position vectors of the magnetic unit cell and the internal sublattice site, respectively:
${\bf R} = \sum_{\mu=1}^d N_{\mu}
{\bf A}_\mu$ with integers $N_{\mu}
\in [0, L)$,
and the number of sublattice sites in a magnetic unit cell is denoted as $N_0 = N/L^d$.
Also, it is useful to define an averaged spin for each sublattice as
\begin{align}
\overline{\bf S}_{{\bf r}_0} =\frac{1}{L^d} \sum_{\bf R} {\bf S}_{{\bf R}+{\bf r}_0}. 
\label{eq:sbar}
\end{align}
Note that $|\overline{\bf S}_{{\bf r}_0}| \leq 1$.
Then, the model in Eq.~\eqref{eq:H_general} can be rewritten in terms of
$\overline{\bf S}_{{\bf r}_0}$ because ${\bf S}_{{\bf Q}_\eta}$ is expressed as 
${\bf S}_{{\bf Q}_\eta} = \sqrt{L^d/N_0} \sum_{{\bf r}_0} \overline{\bf S}_{{\bf r}_0} 
e^{- i{\bf Q}_\eta \cdot {\bf r}_0}$
under the condition in Eq.~\eqref{eq:condQeta}. 
For instance, the Hamiltonian in Eq.~\eqref{eq:H} is rewritten as
\begin{align}
\mathcal{H}
=&-L^d \Bigr[
\frac{1}{N_0} \sum_{({\bf r}_0,\alpha), ({\bf r}_0',{\alpha'}) }  
\overline{ S}^\alpha_{{\bf r}_0}~
\overline{ \mathcal{J}}_{({\bf r}_0,\alpha) ({\bf r}_0',\alpha')}~
\overline{ S}^{\alpha'} _{{\bf r}'_0} \nonumber\\
&~~~~~~~~~
+ \sum_{{\bf r}_0}  {\bf B}  \cdot \overline{\bf S}_{{\bf r}_0} 
\Bigr], 
\label{eq:H_Sbar}
\end{align}
with a $(3N_0) \times (3N_0)$ matrix, $\overline{\mathcal{J}}$, whose elements are defined as 
\begin{align}
\overline{\mathcal{J}}_{({\bf r}_0,\alpha) ({\bf r}_0',\alpha')} =
\sum_{\eta = 1}^{N_Q}
\left[ \mathcal{J}^{\alpha{\alpha'} }_{{\bf Q}_\eta} e^{- i  {\bf Q}_\eta \cdot ({\bf r}_0 -{\bf r}_0')}
+ {\rm c.c.}
\right].
\label{eq:barJ}
\end{align}

\subsection{Method I}\label{sec:method1}
In method I, to compute the partition function 
\begin{align}
Z=\int \Bigr(
\prod_{{\bf R},{\bf r}_0} d{\bf S}_{{\bf R}+{\bf r}_0}\Bigr)
e^{-\beta \mathcal{H}},
\label{eq:Z}
\end{align}
where $\beta$ is the inverse temperature $\beta=1/T$ taking the Boltzmann constant unity ($k_{\rm B}=1$), 
and the integral of ${\bf S}_{{\bf R}+{\bf r}_0}$ is on the surface of the unit sphere in three dimensions,
we apply a $3N_0$-dimensional Gaussian integral to 
the two-spin interaction part by introducing the auxiliary fields $m^\alpha_{{\bf r}_0}$:
\begin{align}
&\exp \Bigr[ 
\frac{\beta L^d}{N_0} 
\sum_{({\bf r}_0,\alpha),({\bf r}'_0,\alpha')} 
\overline{S}^\alpha_{{\bf r}_0} 
\overline{\mathcal{J}}_{({\bf r}_0,\alpha)({\bf r}_0', \alpha')} 
\overline{S}^{\alpha'}_{{\bf r}_0'} 
\Bigr] \nonumber\\
&=
\sqrt{	\frac{N_0^{3N_0}}{(4\pi \beta L^d)^{3N_0} \det  \overline{\mathcal{J}}}} 
\int_{-\infty}^\infty \bigr(\prod_{{\bf r}_0,\alpha} dm^\alpha_{{\bf r}_0}\bigr) \nonumber\\
&\times
\exp \Bigr[ 
-\frac{N_0}{4 \beta L^d} 
\sum_{({\bf r}_0,\alpha),({\bf r}'_0,\alpha')}
m^\alpha_{{\bf r}_0} 
 \overline{\mathcal{J}}^{-1}_{({\bf r}_0,\alpha)({\bf r}_0',\alpha')}
m^{\alpha'}_{{\bf r}_0'} \nonumber\\
&~~~~~
+  \frac{1}{ L^d}  \sum_{{\bf R},{\bf r}_0} {\bf m}_{{\bf r}_0} \cdot  {\bf S}_{{\bf R}+{\bf r}_0}
\Bigr].
\label{eq:gauss}
\end{align}
After rescaling the variable as ${\bf m}_{{\bf r}_0} \to \tilde{\bf m}_{{\bf r}_0} = {\bf m}_{{\bf r}_0} /(\beta L^d)$,
and performing the integrals of individual ${\bf S}_{{\bf R}+{\bf r}_0}$,
we obtain 
\begin{align}
Z=&
\sqrt{\frac{ ( \beta L^d N_0)^{3N_0} }{(4\pi )^{3N_0} \det \overline{\mathcal{J}} }}
\idotsint_{-\infty}^{\infty}
 \bigr( 
 \prod_{({\bf r}_0, \alpha)} d\tilde{m}_{{\bf r}_0}^\alpha \bigr)
 e^{L^d g (\{ \tilde{m}_{{\bf r}_0}^\alpha \})},
\end{align}
where
\begin{align}
g (\{ \tilde{m}_{{\bf r}_0}^\alpha \} )=&
\sum_{{\bf r}_0}
{\rm ln} \left[ 
\frac{4 \pi \sinh ( \beta |\tilde{\bf m}_{{\bf r}_0} + {\bf B} |)}{ \beta |\tilde{\bf m}_{{\bf r}_0} + {\bf B} |}
\right] \nonumber \\
&-
\frac{\beta N_0 }{4}\sum_{({\bf r}_0,\alpha),({\bf r}'_0,\alpha')}
\tilde{m}^\alpha_{{\bf r}_0} 
 \overline{\mathcal{J}}^{-1}_{({\bf r}_0,\alpha)({\bf r}_0',\alpha')}
\tilde{m}^{\alpha'}_{{\bf r}_0'}.
\label{eq:g}
\end{align}
In the thermodynamic limit of $L\to\infty$,
the partition function asymptotically approaches
\begin{align}
Z \to e^{L^d g (\{ \overline{ \tilde{m}_{{\bf r}_0}^\alpha } \})},
\label{eq:Z_saddle}
\end{align}
where
$\{\overline{\tilde{m}^\alpha_{{\bf r}_0}} \}$ denotes the saddle point that maximizes $g (\{ \tilde{m}_{{\bf r}_0}^\alpha\} )$. 
The saddle point is obtained by the stationary condition,
\begin{align}
\frac{ \partial g (\{ \tilde{m}^\alpha_{{\bf r}_0} \})}{\partial \tilde{m}^\alpha_{{\bf r}_0}} =0. 
\end{align}
This leads to a set of equations,
\begin{align}
\tilde{M}^{\alpha'}_\eta =&
\frac{2}{N_0}
\sum_{({\bf r}_0,\alpha)}
  {\mathcal{J}}_{{\bf Q}_\eta}^{\alpha' \alpha}e^{+ i  {\bf Q}_\eta \cdot {\bf r}_0}
\frac{
\tilde{m}^\alpha_{{\bf r}_0} + {B}^\alpha
}{
|\tilde{\bf m}_{{\bf r}_0} + {\bf B}|} \nonumber\\
&\times \left[
\coth ( \beta |\tilde{\bf m}_{{\bf r}_0} + {\bf B} |)
-
\frac{1}{\beta |\tilde{\bf m}_{{\bf r}_0} + {\bf B}|}
\right],\label{eq:M2m}\\
\tilde{m}_{{\bf r}_0}^\alpha =& \sum_{\eta=1}^{N_Q} \Bigr[
\tilde{M}_\eta^\alpha e^{- i  {\bf Q}_\eta \cdot {\bf r}_0}+{\rm c.c.}
\Bigr],
\label{eq:m2M}
\end{align}
which are solved in a self-consistent way.
We will remark on the conditions for the existence of the saddle point solution in Sec.~\ref{subsec:remark}.

Once the saddle point solution is obtained, the free energy per spin is obtained as
\begin{align}
f= - \frac{1}{\beta N} {\rm ln}Z
= - \frac{1}{\beta N_0} g (\{ \overline{ \tilde{m}_{{\bf r}_0}^\alpha } \}).
\end{align}
It is also straightforward to compute other thermodynamic quantities. 
For instance, the internal energy per spin, $\varepsilon$, is obtained by $- T^2 \partial (f/T)/\partial T$. 
The same result is obtained directly from the Hamiltonian in Eq.~\eqref{eq:H} by replacing ${\bf S}_{\bf r}$ with $\langle {\bf S}_{\bf r} \rangle$. 
The specific heat per spin, $C$, is obtained by a numerical derivative of $\varepsilon$ as $C = \partial \varepsilon /\partial T$.
In addition, the real-space spin configuration is obtained 
by
\begin{align}
\langle S^\alpha_{{\bf R}+{\bf r}_0} \rangle 
= \left.
-N_0
\frac{\partial f}{
\partial  B^\alpha_{{\bf r}_0}
} \right|_{{\bf B}_{{\bf r}_0} \to {\bf B}} ,
\end{align}
where ${\bf B}$ in $g(\{\overline{\tilde{m}^\alpha_{{\bf r}_0}} \})$ is replaced by a sublattice dependent field ${\bf B}_{{\bf r}_0}$.
This leads to
\begin{align}
\langle {\bf S}_{{\bf R}+{\bf r}_0} \rangle
=&
\left[ 
\coth ( \beta | \overline{\tilde{\bf m}_{{\bf r}_0}} +{\bf B} | )
-\frac{1}{ \beta  | \overline{\tilde{\bf m}_{{\bf r}_0}} +{\bf B} | }
\right] 
\frac{ \overline{\tilde{\bf m}_{{\bf r}_0}} +{\bf B} }{| \overline{\tilde{\bf m}_{{\bf r}_0}} +{\bf B} |},
\label{eq:m2S}
\end{align}
which is independent of ${\bf R}$; namely, 
$\langle S^\alpha_{{\bf R}+{\bf r}_0} \rangle$ takes the same value at all the sites belonging to the same sublattice.

Finally, let us make a remark on the ground state. 
In Eq.~\eqref{eq:m2S},
the factor in the square brackets becomes unity at zero temperature,
and the spin texture is given by the sum of ${\bf B}$ and the spin density waves or the spin spirals as
\begin{align}
\langle {\bf S}_{{\bf R}+{\bf r}_0} \rangle
= {\cal N}
\Biggl[ 
 \sum_{\eta=1}^{N_Q} \Bigr( \;
\overline{ \tilde{\bf M}_\eta } e^{- i  {\bf Q}_\eta \cdot {\bf r}_0}+{\rm c.c.}
\Bigr)
  +{\bf B}
\Biggr], \label{eq:Sr}
\end{align}
where ${\cal N}$ is the normalization factor to ensure $|\langle {\bf S}_{{\bf R}+{\bf r}_0} \rangle|=1$, and
$\overline{ \tilde{\bf M}_\eta } $ corresponds to the solution of the self-consistent equations in Eqs.~\eqref{eq:M2m} and \eqref{eq:m2M}.
While the expression in Eq.~\eqref{eq:Sr} includes only the Fourier components with ${\bf Q}_\eta$,
$\langle {\bf S}_{{\bf R}+{\bf r}_0} \rangle$ has higher harmonics such as ${\bf S}_{2{\bf Q}_1}$ in general because of the normalization.
This type of spin texture has been discussed for understanding the motion of hedgehogs and antihedgehogs in 3D HLs under the external magnetic field~\cite{Zhang2016,Shimizu2021b}.

\subsection{Method II}\label{sec:method2}
Next, we describe the other method, method II, which is applicable to the generic form of the model in Eq.~\eqref{eq:H_general}.
The key idea of this method is a reduction of the number of integral variables in Eq.~\eqref{eq:Z} by introducing ``density of state''.
Recalling that the Hamiltonian can be written in terms of the averaged spin $\overline{\bf S}_{{\bf r}_0}$ defined in Eq.~\eqref{eq:sbar},  
we can calculate the partition function as
\begin{align}
Z=\int \Bigr[
\prod_{{\bf r}_0} d\overline{\bf S}_{{\bf r}_0}
\rho_{L^d} ( \overline{\bf S}_{{\bf r}_0} )
\Bigr]
e^{-\beta \mathcal{H}},
\label{eq:Z2}
\end{align}
where 
$\int d\overline{\bf S}_{{\bf r}_0}$ denotes an integral inside the unit sphere in three dimensions,
and $\rho_{L^d} ( \overline{\bf S}_{{\bf r}_0} )$ is the density of state for $\overline{\bf S}_{{\bf r}_0}$. 
The number of integral variables is reduced from $2N$ in Eq.~\eqref{eq:Z} to $3N_0$ in Eq.~\eqref{eq:Z2}.

The quantity $\rho_{L^d}(\overline{\bf S}) d\overline{\bf S}/(4\pi)^{L^d}$
represents a probability that the mean vector of 
$L^d$ 3D random vectors uniformly distributed on the unit sphere 
is found in the infinitesimal volume $d\overline{\bf S}$ at $\overline{\bf S}$.
This is equivalent to the Pearson random walk~\cite{Pearson1905,Kiefer1984}. 
From this observation, 
$\rho_{L^d}(\overline{\bf S})$ in the limit of $L^d \to \infty$ 
can be obtained as
\begin{align}
\frac{1}{L^d} {\rm ln} \rho_{L^d} (\overline{\bf S} )
\to {\rm ln} \left[\frac{4\pi \sinh v_0(|\overline{\bf S}|)}{v_0(|\overline{\bf S}|)} \right]
-v_0 (|\overline{\bf S}|) |\overline{\bf S}|, 
\end{align}
where $v_0(|\overline{\bf S}|)$ is determined by numerically solving 
\begin{align}
\coth v_0(|\overline{\bf S}|) -\frac{1}{v_0(|\overline{\bf S}|)} = |\overline{\bf S}|.
\end{align}
Using this form, we obtain an asymptotic form of the partition function in the thermodynamic limit as
\begin{align}
Z \to \int \bigr(
\prod_{{\bf r}_0} d\overline{\bf S}_{{\bf r}_0}
\bigr)
e^{L^d G(\{ \overline{S}_{{\bf r}_0}^\alpha \})}, 
\label{eq:Zasym}
\end{align}
where
\begin{align}
&
G(\{ \overline{S}_{{\bf r}_0}^\alpha \})= 
\nonumber\\
&~
-\frac{\beta}{L^d}
 \mathcal{H} + 
\sum_{{\bf r}_0} \Bigr[
{\rm ln} \Bigr[\frac{4\pi \sinh v_0(|\overline{\bf S}_{{\bf r}_0} |)}{v_0(|\overline{\bf S}_{{\bf r}_0}|)} \Bigr]
-v_0 (|\overline{\bf S}_{{\bf r}_0}|) |\overline{\bf S}_{{\bf r}_0}|
\Bigr].
\end{align}
Then, by the steepest descent method, 
the partition function is expressed as
\begin{align}
Z
\sim
 e^{L^d G (\{ \overline{ \overline{S}_{{\bf r}_0}^\alpha } \})},
\label{eq:Z_saddle2}
\end{align}
where
$\{ \overline{ \overline{S}_{{\bf r}_0}^\alpha } \}$ denotes the saddle point that maximizes $G (\{ \overline{S}_{{\bf r}_0}^\alpha \} )$. 
In comparison with Eq.~\eqref{eq:Z_saddle} in method I, we note that 
$G(\{ \overline{ \overline{S}_{{\bf r}_0}^\alpha} \} ) = g (\{ \overline{ \tilde{m}_{{\bf r}_0}^\alpha } \})$.
Once the saddle point solution is obtained, the thermodynamic quantities and the real-space spin configurations are computed in a similar manner to method I.

\subsection{Remark}\label{subsec:remark}
As both method I and II are based on the steepest descent method, the saddle point solution exists only when $g (\{ \overline{ \tilde{m}_{{\bf r}_0}^\alpha } \})$ in Eq.~\eqref{eq:Z_saddle} and $G(\{ \overline{ \overline{S}_{{\bf r}_0}^\alpha} \} )$ in Eq.~\eqref{eq:Z_saddle2} have maxima in the parameter space. 
This is guaranteed when the Hessian matrices of $-g (\{ \tilde{m}_{{\bf r}_0}^\alpha \})$ and $-G(\{ \overline{S}_{{\bf r}_0}^\alpha \})$ are positive definite 
at $\{ \overline{ \tilde{m}_{{\bf r}_0}^\alpha } \}$ and $\{\overline{ \overline{S}_{{\bf r}_0}^\alpha}\}$, respectively.
In method I, this corresponds to the condition that $\overline{\mathcal{J}}$ is positive definite. 
In this case, however, since $\overline{\mathcal{J}}$ is made of the Fourier components of $\pm {\bf Q}_\eta$ only [Eq.~\eqref{eq:barJ}], 
the $6N_Q$ eigenvalues are given by those of $\mathcal{J}_{\pm {\bf Q}_\eta}$, while the rest $3N_0 - 6 N_Q$ eigenvalues are all zero. 
To avoid such zero eigenvalues, we add a positive infinitesimal $\lambda$ to all the diagonal elements of $\overline{\mathcal{J}}$, namely,
$ \overline{\mathcal{J}} \to \overline{\mathcal{J}}+\lambda \mathcal{I}$ with an identity matrix $\mathcal{I}$, 
and take the limit of $\lambda \to 0^+$ in the end of the calculations.
This consideration ensures that $\overline{ \tilde{m}_{{\bf r}_0}^\alpha }$ in Eq.~\eqref{eq:m2M} includes the Fourier components of $\pm {\bf Q}_\eta$ only because $g (\{ \overline{ \tilde{m}_{{\bf r}_0}^\alpha } \})$ negatively diverges due to the second term of Eq.~\eqref{eq:g} as $ - \mathcal{O}(1/\lambda) \xrightarrow[]{\lambda \to 0^+} -\infty$ 
when $\overline{ \tilde{m}_{{\bf r}_0}^\alpha }$ includes the Fourier components of ${\bf q} \neq \pm {\bf Q}_\eta$.

Method I is computationally cheaper than method II 
in most cases, since the number of variables to be determined in method I ($3N_Q$) is typically less than that in method II ($3N_0$). 
It is applicable to the model with two-spin interactions in Eq.~\eqref{eq:H}  as long as $\mathcal{J}_{{\bf Q}_\eta}$ is positive definite as discussed above.
Meanwhile, method II has a wider range of applications. 
In this case, $\mathcal{J}_{{\bf Q}_\eta}$ in the two-spin interaction part does not have to be positive definite.
Furthermore, method II can deal with the generic form of the Hamiltonian expressed by a function of $S^\alpha_{{\bf Q}_\eta}$, including multiple-spin interactions,
such as the biquadratic ones $\propto ({\bf S}_{{\bf Q}_\eta} \cdot {\bf S}_{-{\bf Q}_\eta})^2$ ~\cite{Hayami2017,Hayami2020,Hayami2020b,Okumura2020,Yasui2020,Hayami2021d,Hayami2021,Hayami2021f,Hayami2021g,Hayami2021h,Hirschberger2021,Seo2021}
and the bicubic ones 
$\propto \{
{\bf S}_{{\bf Q}_1} \cdot (
{\bf S}_{{\bf Q}_2}
\times
{\bf S}_{{\bf Q}_3})
\}^2$~\cite{Hayami2021c}.

\section{Results}\label{sec:results}

In this section, as a demonstration of our framework developed in Sec.~\ref{sec:method}, 
we study two models, both of which are in the class of the models with only two-spin interactions, 
as represented by Eqs.~\eqref{eq:H} and \eqref{eq:Hq2}.
Specifically, for both models, we consider
\begin{align}
\mathcal{H}_{{\bf Q}_\eta} =& -
\sum_{\alpha, \alpha' } J^{\alpha \alpha' }_{\eta} S^\alpha_{{\bf Q}_\eta} S^{\alpha'} _{-{\bf Q}_\eta} 
- i  {\bf D}_{\eta} \cdot ({\bf S}_{{\bf Q}_\eta} \times {\bf S}_{-{\bf Q}_\eta} ), \label{eq:Hq}
\end{align}
where the first term represents the symmetric exchange interactions 
with $J_{\eta}^{\alpha\alpha'}=J_{\eta}^{\alpha'\alpha} \in \mathbb{R}$, 
and the second term represents the antisymmetric ones of the DM type~\cite{Dzyaloshinsky1958,Moriya1960}
with the so-called DM vectors ${\bf D}_{\eta}=(D^x_\eta,D^y_\eta,D^z_\eta) \in \mathbb{R}^3$. 
For simplicity, 
we assume that the symmetric interactions include only the diagonal elements, 
and that the DM vectors are proportional to the corresponding characteristic wavenumber:
\begin{align}
J_{\eta}^{\alpha \alpha'} = J_{\eta}^{\alpha\alpha} \delta_{\alpha \alpha'},
\quad
{\bf D}_\eta = D \frac{{\bf Q}_\eta}{|{\bf Q}_\eta|}, 
\label{eq:JD}
\end{align}
where $\delta_{\alpha \alpha'}$ is the Kronecker delta.
Note that the latter assumption leads the system to prefer proper-screw type magnetic orders.
Then, the Hermitian matrix $\mathcal{J}_{{\bf Q}_\eta}$ in Eq.~\eqref{eq:Hq2} is expressed as
\begin{align}
\mathcal{J}_{{\bf Q}_\eta} =
\begin{bmatrix}
J^{xx}_{\eta} & i D^z_{\eta} & - i D^y_{\eta} \\
-i D^z_{\eta} & J^{yy}_{\eta} & i D^x_{\eta} \\
i D^y_{\eta} & -i D^x_{\eta} & J^{zz}_{\eta}
\end{bmatrix}
.
\label{eq:J_q_matrix}
\end{align} 
We define the two models on a simple cubic lattice ($d=3$) with the lattice constant being unity under the periodic boundary condition.
In the following, we set the elements of the commensurate wave numbers ${\bf Q}_\eta$ as $\pm2\pi/\Lambda$ or $0$ with an integer $\Lambda$. 
In this setting, the magnetic unit cell fits into a cube of $\Lambda^3$ sites with the magnetic translation vectors ${\bf A}_1=(\Lambda,0,0)$,
${\bf A}_2=(0,\Lambda,0)$, and
${\bf A}_3=(0,0,\Lambda)$, namely, 
$l_{\mu\mu'}$ in the equation for ${\bf A}_\mu$ above Eq.~\eqref{eq:condQeta} is $l_{\mu\mu'}=\Lambda \delta_{\mu \mu'}$.
Then, the linear dimension of the entire system is $L \Lambda$, 
the lattice site ${\bf r}$ is denoted as ${\bf r}= (x, y, z)$ with integers $x$, $y$, and $z$ in $[0,L\Lambda)$, 
and the number of spin is $N = (L \Lambda)^3$. 
In the following calculations, we take $\Lambda = 12$.

The difference between the two models lies in the number of the characteristic wave numbers ${\bf Q}_\eta$. 
One of them has three ($\eta=1, 2, 3$), and the other has six ($\eta=1, 2, \ldots, 6$). 
We call the former the $3Q$ model and the latter the $6Q$ model. 
The directions of ${\bf Q}_\eta$ as well as the form of $J_\eta^{\alpha\alpha}$ are defined in the following subsections.
We present the results for the $3Q$ and $6Q$ models in Secs.~\ref{subsec:3Q} and \ref{subsec:6Q}, respectively. 
For both cases, we clarify the ground-state phase diagram at zero magnetic field while changing the interaction parameters, 
and the finite-temperature phase diagram in a magnetic field for representative sets of the interaction parameters. 
Finally, in Sec.~\ref{sec:hidden_transition}, we give some remarks on hidden transitions found through detailed analyses.

All the results in this section are obtained by method I in Sec.~\ref{sec:method1}, 
while we confirm that method II in Sec.~\ref{sec:method2} delivers the same result for some parameter values. 
In what follows, we omit the overline of $\overline{\tilde{\bf M}_\eta}$ for simplicity and use $\tilde{\bf M}_\eta$ to represent the solution of the self-consistent equations in Eqs.~\eqref{eq:M2m} and \eqref{eq:m2M}.

\subsection{$3Q$ model}\label{subsec:3Q}
First, we discuss the model with three ${\bf Q}_\eta$, the $3Q$ model. 
After introducing the model parameters in Sec.~\ref{sec:modelparam3Q}, 
we present the ground-state phase diagram at zero magnetic field while varying the anisotropy in the symmetric interaction, $\Delta$, and the magnitude of the DM vectors, $D$, in Sec.~\ref{sec:GroundState3Q}.
Then, in Sec.~\ref{sec:FiniteT3Q}, we show the magnetic field-temperature phase diagrams for a couple of representative parameter sets of $\Delta$ and $D$.

\subsubsection{Model parameters}\label{sec:modelparam3Q}
\begin{figure}[bhtp]
  \centering
  \includegraphics[trim=0 0 0 0, clip,width=\columnwidth]{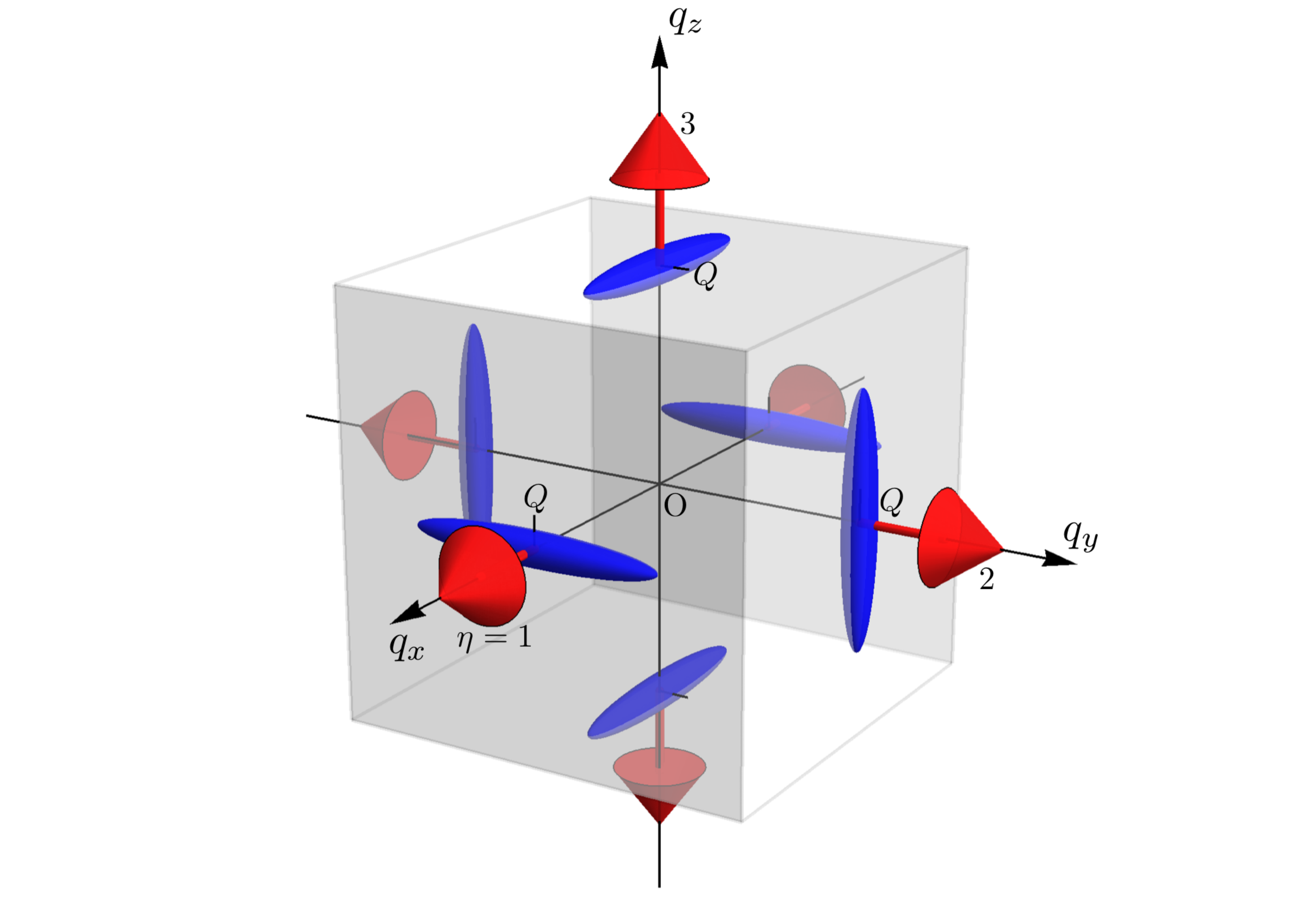}
  \caption{
  Pictorial representation of the coupling constants for the symmetric and antisymmetric interactions in the $3Q$ model.
  The blue ellipsoids at $\pm{\bf Q}_\eta$ represent $J^{\alpha\alpha}_{\eta}$: The lengths along the principal axes [100], [010], and [001] denote the amplitudes of $J^{xx}_{\eta}$, $J^{yy}_{\eta}$, and $J^{zz}_{\eta}$, respectively. The red arrows at $\pm{\bf Q}_\eta$ represent $\pm{\bf D}_{\eta}$.
   The labeled numbers indicate $\eta$.
   The gray cube is a guide to the eye.
  }
  \label{fig01}
\end{figure}
The model Hamiltonian is given by Eqs.~\eqref{eq:H} and \eqref{eq:Hq} with $N_Q=3$. 
We set ${\bf Q}_\eta$ as
\begin{align}
{\bf Q}_1 = Q \hat{\bf x},~
{\bf Q}_2 = Q \hat{\bf y},~
{\bf Q}_3 = Q \hat{\bf z},
\label{eq:Q3Q}
\end{align}
where $Q=2\pi/\Lambda$ with $\Lambda = 12$, and
$\hat{\bf x}$, $\hat{\bf y}$, and $\hat{\bf z}$ represent unit vectors as $\hat{\bf x}=(1,0,0)$, $\hat{\bf y}=(0,1,0)$, and $\hat{\bf z}=(0,0,1)$.
We introduce an anisotropy $\Delta$ to the symmetric interactions $J_\eta^{\alpha \alpha}$ in Eq.~\eqref{eq:JD} as
\begin{align}
&(J_\eta^{xx},J_\eta^{yy},J_\eta^{zz}) = \nonumber\\
&\begin{cases}
 [J (1-\Delta), J(1+2\Delta), J(1-\Delta)],&(\eta=1)\\
 [J(1-\Delta), J (1-\Delta), J(1+2\Delta)],&(\eta=2)\\
 [ J(1+2\Delta), J(1-\Delta),J (1-\Delta)],&(\eta=3)\\
 \end{cases}.~\label{eq:J3Q}
\end{align}
As shown in Sec.~\ref{sec:GroundState3Q}, the anisotropy stabilizes a $3Q$ magnetic order.
We take the DM vectors in the antisymmetric interaction as ${\bf D}_\eta \parallel {\bf Q}_\eta$ as in Eq.~\eqref{eq:JD}. 
Figure~\ref{fig01} shows the pictorial representation of $J_\eta^{\alpha \alpha}$ and ${\bf D}_\eta$.
We take the energy unit as $J=1$.

\subsubsection{Ground state at zero magnetic field}\label{sec:GroundState3Q}
\begin{figure}[bhtp]
  \centering
  \includegraphics[trim=0 0 0 0, clip,width=\columnwidth]{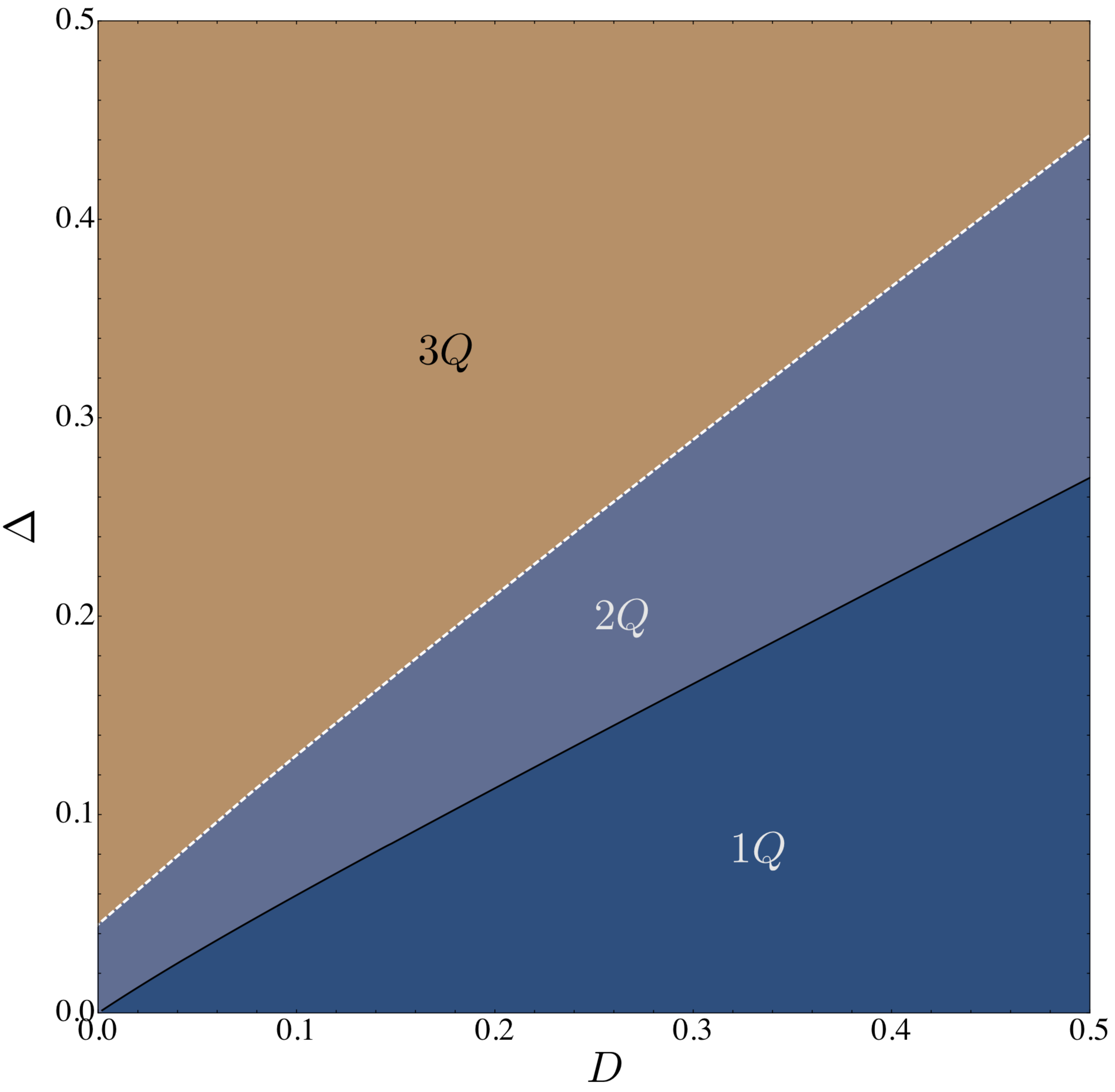}
  \caption{
  Ground-state phase diagram for the $3Q$ model at zero magnetic field.
  The phase diagram includes the $1Q$ phase where one of three $|\tilde{\bf M}_{\eta}|$ is nonzero, 
  the $2Q$ phase where two of $|\tilde{\bf M}_{\eta}|$ are nonzero at different values, 
  and the isotropic $3Q$ phase where $|\tilde{\bf M}_1| = |\tilde{\bf M}_2| = |\tilde{\bf M}_3|>0$.
  The white dashed line between the $3Q$ and $2Q$ phases represents the first-order phase transition, 
  while the black solid line between the $2Q$ and $1Q$ phases represents the second-order phase transition.
  }
  \label{fig02}
\end{figure}
\begin{figure}[bhtp]
  \centering
  \includegraphics[trim=0 0 0 0, clip,width=\columnwidth]{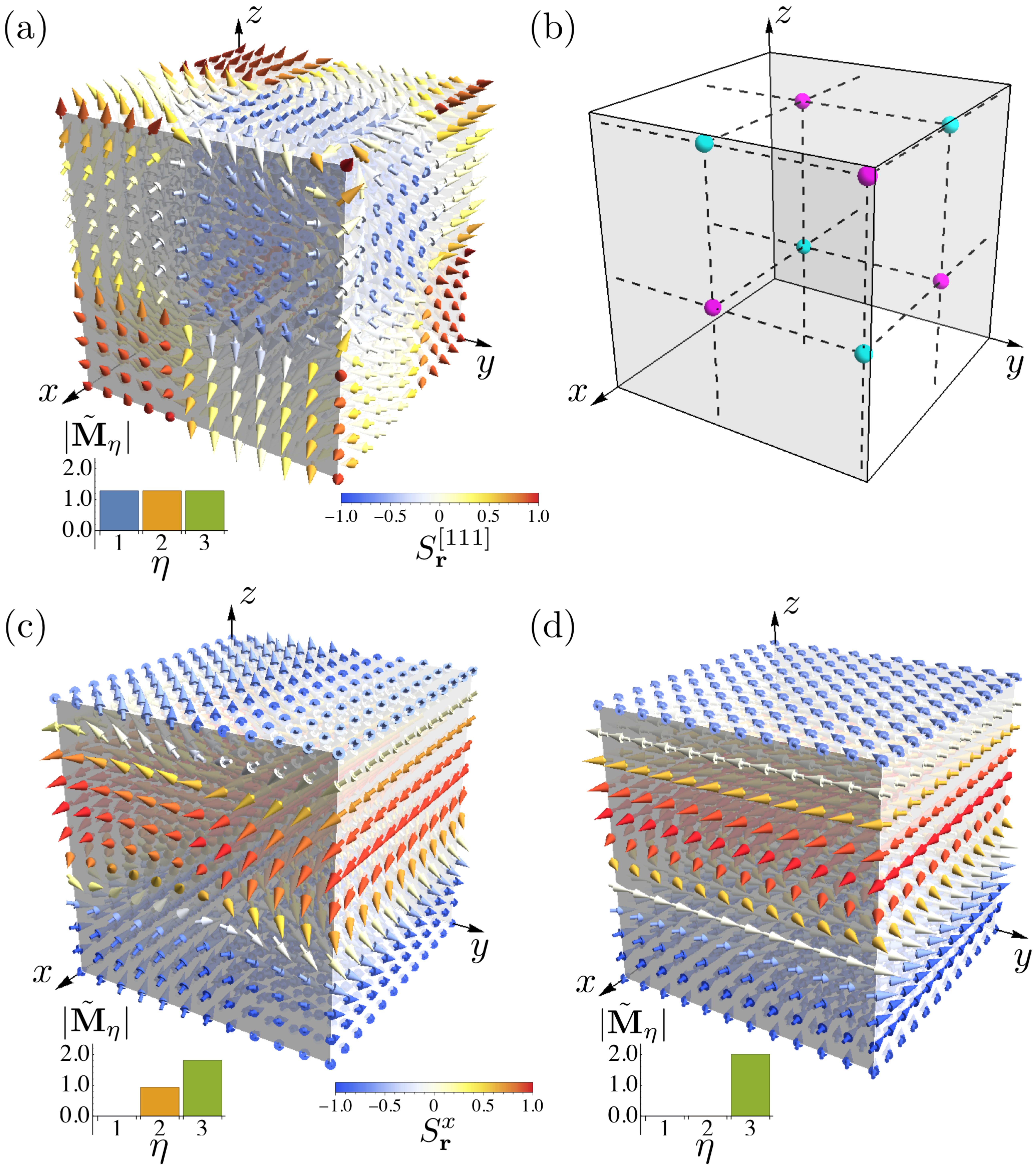}
  \caption{
   Ground-state spin configurations stabilized in the $3Q$ model at zero magnetic field for
   (a) the $3Q$ state at $(D,\Delta)=(0.15,0.3)$,
   (c) the $2Q$ state at $(D,\Delta)=(0.25,0.2)$, and
   (d) the $1Q$ state at $(D,\Delta)=(0.35,0.1)$.
   The color of the arrows denotes the $[111]$ component of spins, $S^{[111]}_{\bf r} = (S^x_{\bf r} + S^y_{\bf r} + S^z_{\bf r})/\sqrt{3}$, in (a), 
   while it represents the $x$ component of spins, $S^x_{\bf r}$, in (c) and (d); 
   see the color bars in (a) and (c).
   (b) Positions of the magnetic hedgehogs (magenta spheres) and the magnetic antihedgehogs (cyan spheres) in the $3Q$ spin state in (a). 
   The dashed lines are guides to the eye.
   Insets of (a), (c), and (d) show distributions of $|\tilde{\bf M}_\eta|$ for each spin state. 
  }
  \label{fig03}
\end{figure}
\begin{figure*}[htbp]
  \centering
  \includegraphics[trim=0 0 0 0, clip,width=\textwidth]{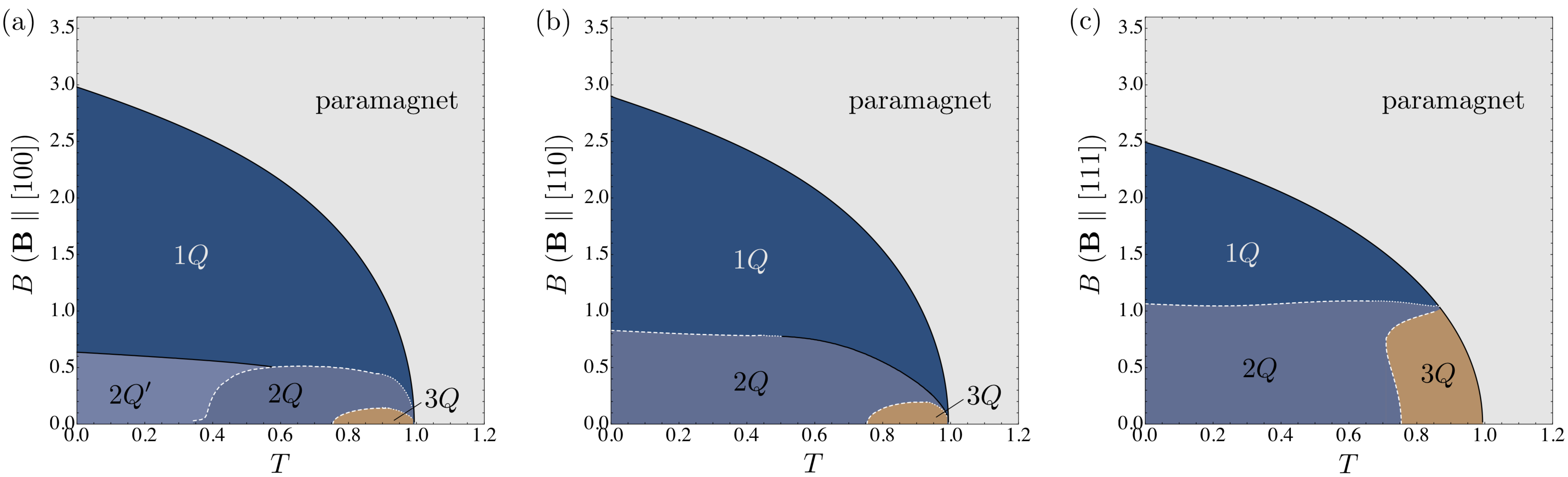}
  \caption{
  Magnetic field-temperature phase diagrams of the $3Q$ model with $(D,\Delta)=(0.25,0.2)$ 
  for the magnetic field directions
  (a) ${\bf B}\parallel [100]$,
  (b) ${\bf B}\parallel [110]$, and
  (c) ${\bf B}\parallel [111]$.
  The white dashed and black solid lines represent first-order and second-order phase transitions, respectively, 
  while the white dotted lines represent phase transitions whose order is undetermined.
  }
  \label{fig04}
\end{figure*}
\begin{figure*}[htbp]
  \centering
  \includegraphics[trim=0 0 0 0, clip,width=\textwidth]{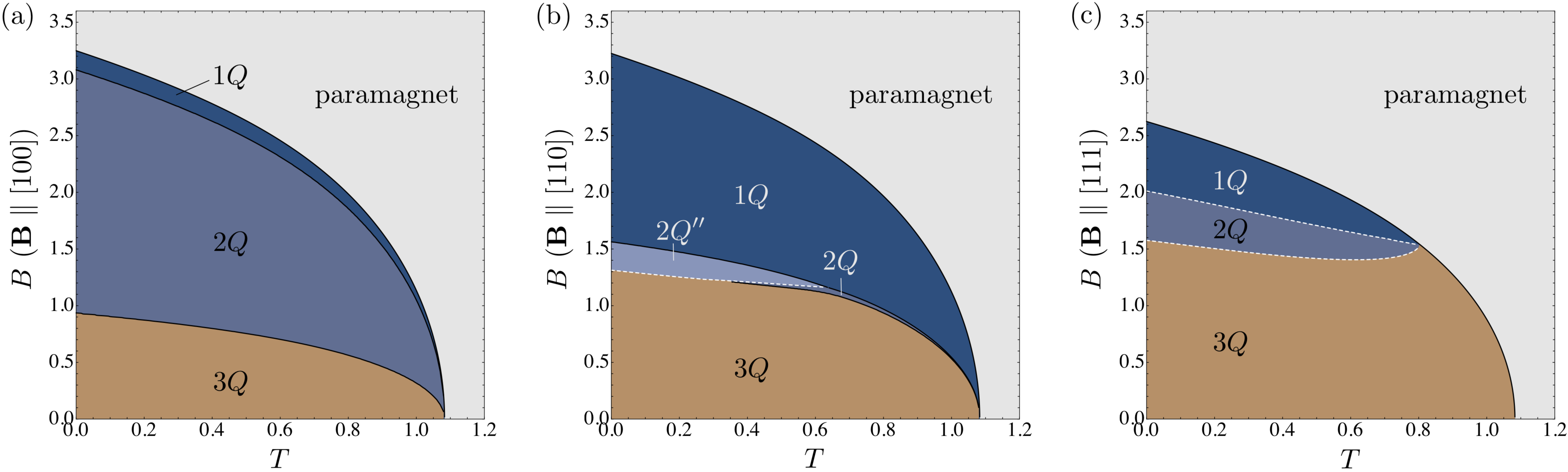}
  \caption{
  Magnetic field-temperature phase diagrams of the $3Q$ model with $(D,\Delta)=(0.15,0.3)$ 
  for the magnetic field directions
  (a) ${\bf B}\parallel [100]$,
  (b) ${\bf B}\parallel [110]$, and
  (c) ${\bf B}\parallel [111]$.
  The notations are common to those in Fig.~\ref{fig04}. 
  }
  \label{fig05}
\end{figure*}
\begin{figure}[htpb]
  \centering
  \includegraphics[trim=0 0 0 0, clip,width=\columnwidth]{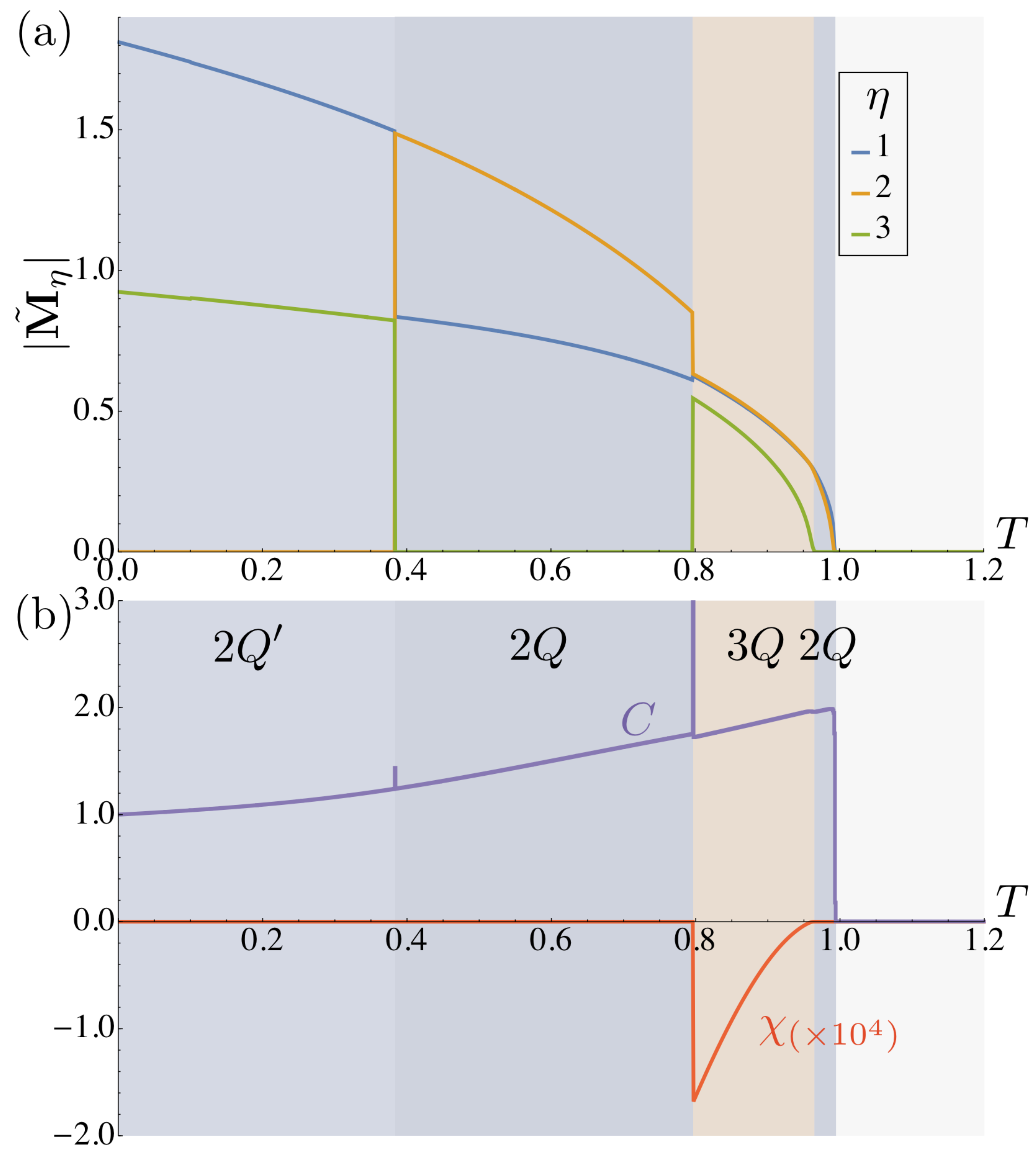}
  \caption{
  Temperature dependences of 
  (a) the order parameters $|\tilde{\bf M}_\eta|$ and (b) the specific heat $C$ and the spin scalar chirality $\chi$
  for the $3Q$ model with $(D,\Delta)=(0.25,0.2)$ and $B=0.1$ (${\bf B}\parallel [100]$).
  The spin scalar chirality is multiplied by a factor of $10^4$ for better visibility.
  }
  \label{fig06}
\end{figure}
\begin{figure*}[htpb]
  \centering
  \includegraphics[trim=0 0 0 0, clip,width=\textwidth]{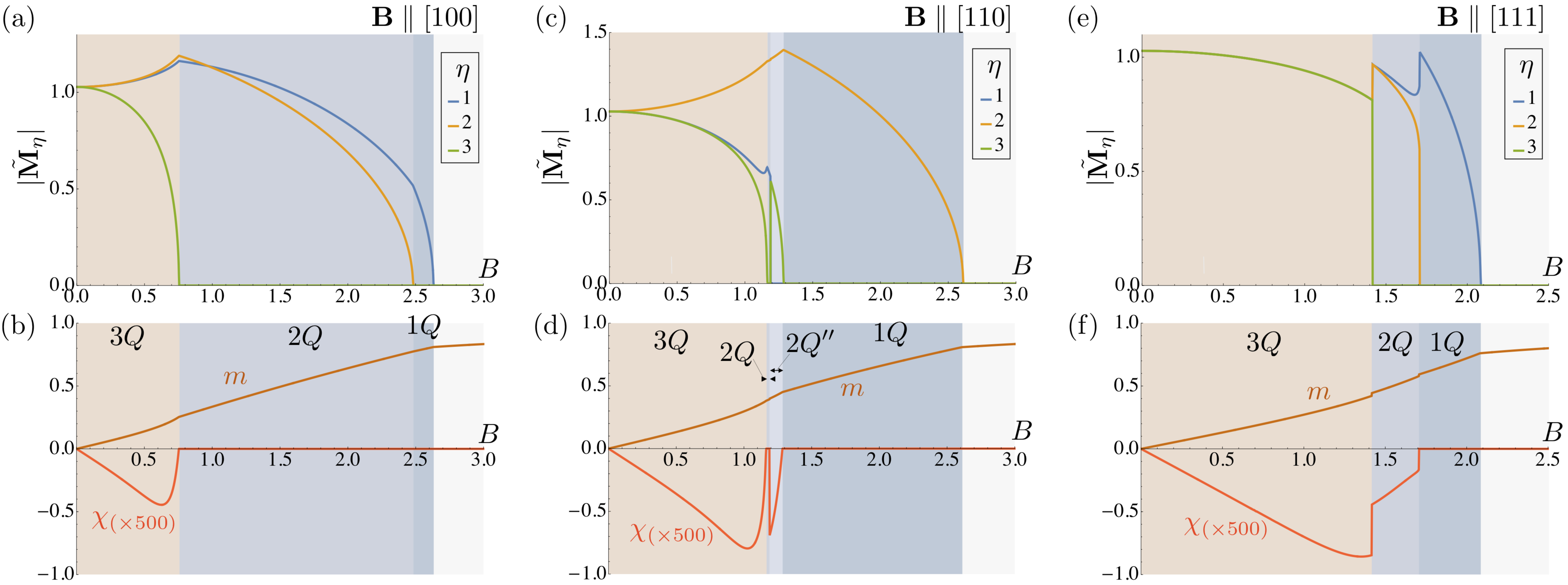}
  \caption{
  Magnetic field dependences of 
  (a,c,e) the order parameters $|\tilde{\bf M}_\eta|$ and  (b,d,f) the magnetization $m$ and the spin scalar chirality $\chi$
  for the $3Q$ model with $(D,\Delta)=(0.15,0.3)$ and $T=0.5$. 
  The magnetic field directions are 
   (a,b) ${\bf B}\parallel [100]$, 
   (c,d) ${\bf B}\parallel [110]$, 
   and
   (e,f) ${\bf B}\parallel [111]$. 
   The spin scalar chirality is multiplied by a factor of $500$, respectively, for better visibility.
  }
  \label{fig07}
\end{figure*}
Figure~\ref{fig02} shows the ground-state phase diagram for the $3Q$ model at zero magnetic field while changing $\Delta$ and $D$. 
The phase diagram with low resolution was obtained by variational calculations in Ref.~\cite{Kato2021}; much higher resolution can be reached here with much less computational cost owing to the use of the present framework. 
As in the previous study, we find three stable phases in the phase diagram:
The $1Q$ phase in the small $\Delta$ region 
including the isotropic limit ($\Delta=0$)~\cite{Nussinov2001}, the $3Q$ phase in the large $\Delta$ region, 
and the $2Q$ phase in between them. 
The phase transition between the $1Q$ and $2Q$ phases is continuous (second order), while that between the $2Q$ and $3Q$ phases is discontinuous (first order). 
If we look into more detail, however, we find a discontinuous transition line in the $3Q$ phase (not shown in the phase diagram); 
we will discuss the hidden transition in Sec.~\ref{sec:hidden_transition}.

We display typical spin textures in the three phases in Fig.~\ref{fig03}, with the values of $|\tilde{\bf M}_\eta|$ in each inset.
Figure~\ref{fig03}(a) represents the $3Q$ state. 
This is a 3D HL that possesses topological point defects, the magnetic hedgehogs and antihedgehogs, forming a periodic lattice, as shown in Fig.~\ref{fig03}(b). 
In this phase, the relation $|\tilde{\bf M}_1| =  |\tilde{\bf M}_2| = |\tilde{\bf M}_3|$ always holds; namely, the $3Q$ state is composed of a superposition of three proper screws with equal amplitudes. 
Meanwhile, Figs.~\ref{fig03}(c) and \ref{fig03}(d) represent the $2Q$ and $1Q$ states, respectively. 
The $2Q$ state is composed of a superposition of two proper screws with different amplitudes in general.
Note that the $\mathsf{C}_3$ rotational symmetry about the $[111]$ axis is retained in the $3Q$ phase, 
whereas it is broken in the $2Q$ and $1Q$ phases.

\subsubsection{Magnetic field-temperature phase diagrams}\label{sec:FiniteT3Q}
Figures~\ref{fig04} and \ref{fig05} show the magnetic field-temperature phase diagrams for the representative parameter sets 
that realize the $2Q$ and $3Q$ ground states at zero magnetic field, respectively.
We take $(D,\Delta)=(0.25,0.2)$ for the $2Q$ case and $(D,\Delta)=(0.15,0.3)$ for the $3Q$ case, for which the ground-state spin configurations at zero magnetic field are shown in Figs.~\ref{fig03}(c) and \ref{fig03}(a), respectively.
In each case, we obtain the results for different magnetic field directions, ${\bf B}\parallel [100]$, ${\bf B}\parallel [110]$, and ${\bf B}\parallel [111]$ in panels (a), (b), and (c), respectively, of Figs.~\ref{fig04} and \ref{fig05} ($B=|{\bf B}|$).

Let us begin with the results for $(D,\Delta)=(0.25,0.2)$ in Fig.~\ref{fig04}, 
where the ground state at zero magnetic field is in the $2Q$ phase.
While increasing temperature at zero field, 
we find two phase transitions:
A first-order phase transition from the $2Q$ to $3Q$ phase at $T\simeq 0.754$ 
and a second-order phase transition from the $3Q$ phase to the paramagnet at $T\simeq 0.994$. 
When we apply the magnetic field, 
regardless of its direction, the $2Q$ and $3Q$ phases are stable in the low field region,  
whereas the $1Q$ phase appears for higher fields.
Types of the transition to the $1Q$ phase depend on the field direction. 
For ${\bf B}\parallel [100]$ in Fig.~\ref{fig04}(a), the transition is of first order in most of the high $T$ regime where the system changes directly from the $2Q$ to $1Q$ phase
although the discontinuity becomes very weak and the order of the transition becomes unclear for $T \gtrsim 0.90$ as indicated by the white dotted line.
Meanwhile, 
there appears an intermediate $2Q'$ phase between the $2Q$ and $1Q$ phases in the low $T$ regime, 
which is a double-$Q$ phase different from the $2Q$ phase (see below). 
The transition from the $2Q$ to $2Q'$ phase and that from the $2Q'$ and $1Q$ phase are of first and second order, respectively.
In contrast, for ${\bf B}\parallel [110]$ in Fig.~\ref{fig04}(b),
the transition to the $1Q$ phase always takes place from the $2Q$ phase, 
but the nature of the transition changes with temperature: 
It is of second order in the high $T$ regime, 
while that becomes first order in the low $T$ regime. 
This suggests the presence of the tricritical point where the two types of the transition lines meet, but it is hard to determine its precise location within the present resolution; it would be located at some point on the white dotted line for $
0.45 \lesssim T\lesssim 0.51$ on which the order of the transition is not precisely determined.
Similarly, for ${\bf B}\parallel [111]$ in Fig.~\ref{fig04}(c), 
the first-order phase transition between the $2Q$ and $1Q$ phases becomes obscure while increasing $T$, 
and the order of the transition is not clear in the high $T$ regime for $T\gtrsim 0.67$.

Interestingly, in all the cases, the $3Q$ phase appears in a domelike shape at finite temperature under the magnetic field. 
It is surrounded by the $2Q$ phase in the cases of ${\bf B} \parallel [100]$ and ${\bf B}\parallel [110]$,
while it borders both the $2Q$ and paramagnetic phases for ${\bf B} \parallel [111]$.
The transition between the $3Q$ and $2Q$ phases is always of first order, while that to the paramagnet is of second order.
The results indicate that the higher multiple-$Q$ phase is induced from the lower one by the entropic gain.

Figure~\ref{fig06}(a) shows the temperature dependence of the order parameters $|\tilde{\bf M}_\eta|$ at a low field of ${\bf B}\parallel [100]$, $B=0.1$, 
where we find successive transitions as $2Q' \to 2Q \to 3Q \to 2Q \to$ paramagnet while increasing temperature. 
At low temperature, a first-order phase transition separates two double-$Q$ phases, the $2Q'$ and $2Q$ phases,
although they are indistinguishable at ${\bf B}=0$.
In both phases, $|\tilde{\bf M}_1|$, which is the component along the magnetic field direction, is nonzero, but it discontinuously changes at the transition. 
The other nonzero component is switched at the transition; 
$|\tilde{\bf M}_3|$ is nonzero in the low-$T$ $2Q'$ phase, while $|\tilde{\bf M}_2|$ becomes nonzero in the high-$T$ $2Q$ phase. 
Note that $|\tilde{\bf M}_2|$ and $|\tilde{\bf M}_3|$ are both perpendicular components to the magnetic field, 
but they are not equivalent due to the anisotropy in the symmetric exchange interactions $J_\eta^{\alpha\alpha}$ (see Fig.~\ref{fig01}). 
In the $3Q$ phase at higher temperature, all $|\tilde{\bf M}_\eta|$ are nonzero; $|\tilde{\bf M}_1|$ and $|\tilde{\bf M}_2|$ take almost the same value, while $|\tilde{\bf M}_3|$ is smaller. 
At the transition to the $2Q$ phase, $|\tilde{\bf M}_3|$ goes to zero continuously, and finally, the remaining two components become zero continuously at the transition to the paramagnet; 
it is unclear whether the two components vanish simultaneously or not in the present resolution, namely whether the $1Q$ state exists or not before entering the paramagnetic phase.
We summarize the order parameters in each phase in Table~\ref{tab1}.

\begin{table}[hbpt]
\caption{\label{tab1}
Order parameters in each magnetically ordered phase of the $3Q$ model. 
The sets of $\eta$, $\{\eta_1,\eta_2,\cdots\}$, 
for nonzero $\tilde{\bf M}_\eta$ are shown, with the relations between nonzero $|\tilde{\bf M}_\eta|$.
}
\begin{ruledtabular}
\begin{tabular}{lccl}
                   & Phase  & Sets of $\eta$ & Notes \\
\hline
${\bf B}=0$ & $1Q$ & $\{1\}$, $\{2\}$, $\{3\}$ &\\
		  & $2Q$ & $\{3,2\}$, $\{2,1\}$, $\{1,3\}$ & $|\tilde{\bf M}_{\eta_1} | > |\tilde{\bf M}_{\eta_2} |$\\ 
                   & $3Q$ & $\{1,2,3\}$ & $|\tilde{\bf M}_{\eta_1} | = |\tilde{\bf M}_{\eta_2} |= |\tilde{\bf M}_{\eta_3} |$
 \vspace{0.1cm}\\
\hline
${\bf B}\parallel [100]$ & $1Q$ & $\{1\}$ &\\
                   & $2Q$ & $\{1,2\}$ & $|\tilde{\bf M}_{\eta_1} | \neq |\tilde{\bf M}_{\eta_2} |$ \\ 
                   & $2Q'$ & $\{1,3\}$ & $|\tilde{\bf M}_{\eta_1} | > |\tilde{\bf M}_{\eta_2} |$\\ 
                   & $3Q$ & $\{2,1,3\}$ & $|\tilde{\bf M}_{\eta_1} | > |\tilde{\bf M}_{\eta_2} | > |\tilde{\bf M}_{\eta_3} | $
 \vspace{0.1cm}\\
\hline
${\bf B}\parallel [110]$ & $1Q$  & $\{2\}$ &\\
                   & $2Q$ & $\{2,1\}$  & $|\tilde{\bf M}_{\eta_1} | > |\tilde{\bf M}_{\eta_2} |$\\ 
                   & $2Q''$ & $\{2,3\}$ & $|\tilde{\bf M}_{\eta_1} | > |\tilde{\bf M}_{\eta_2} |$\\ 
                   & $3Q$ & $\{2,1,3\}$ & $|\tilde{\bf M}_{\eta_1} | > |\tilde{\bf M}_{\eta_2} | > |\tilde{\bf M}_{\eta_3} | $ 
 \vspace{0.1cm}\\
\hline
${\bf B}\parallel [111]$ & $1Q$ & $\{1\}$, $\{2\}$, $\{3\}$, &\\
                   & $2Q$ & $\{3,2\}$, $\{2,1\}$, $\{1,3\}$ & $|\tilde{\bf M}_{\eta_1} | > |\tilde{\bf M}_{\eta_2} |$\\ 
                   & $3Q$ & $\{1,2,3\}$ & $|\tilde{\bf M}_{\eta_1} | = |\tilde{\bf M}_{\eta_2} |= |\tilde{\bf M}_{\eta_3} |$
                   \\
\end{tabular}
\end{ruledtabular}
\end{table}

Figure~\ref{fig06}(b) shows the temperature dependences of 
the specific heat per spin, $C$, and the spin scalar chirality per spin, $\chi$. 
The latter is defined as 
\begin{align}
\chi =& \frac{1}{\Lambda^3} \sum_{{\bf r}_0} {\boldsymbol \chi}_{{\bf r}_0}
\cdot \hat{\bf e}_{{\bf B}},
\label{eq:chi}
\end{align}
with
\begin{align}
{\chi}^\mu_{{\bf r}_0} =&
\frac{1}{2} 
\bigr[
\langle {\bf S}_{{\bf r}_0} \rangle \cdot \bigr(
\langle {\bf S}_{{\bf r}_0+\hat{\boldsymbol \nu}} \rangle \times 
\langle {\bf S}_{{\bf r}_0+\hat{\boldsymbol \gamma}}  \rangle+
\langle {\bf S}_{{\bf r}_0+\hat{\boldsymbol \gamma}} \rangle \times 
\langle {\bf S}_{{\bf r}_0-\hat{\boldsymbol \nu}} \rangle \nonumber\\
&+
\langle {\bf S}_{{\bf r}_0-\hat{\boldsymbol \nu}} \rangle \times 
\langle {\bf S}_{{\bf r}_0-\hat{\boldsymbol \gamma}} \rangle
+
\langle {\bf S}_{{\bf r}_0-\hat{\boldsymbol \gamma}} \rangle \times 
\langle {\bf S}_{{\bf r}_0+\hat{\boldsymbol \nu}} \rangle
\bigr)
\bigr],
\end{align}
where $\{\mu,\nu,\gamma\} = \{x,y,z\}$, $\{y,z,x\}$, or $\{z,x,y\}$,
and $\hat{\bf e}_{{\bf B}} = {\bf B}/|{\bf B}|$.
The specific heat shows clear anomalies at the transition between the $2Q$ and $3Q$ phases and that to the paramagnet: 
The former is a delta-function type anomaly characteristic to the first-order transition, 
and the latter shows a jump similar to the second-order phase transition in the mean-field approximation.
In contrast, $C$ shows less anomalies at the transition between the $2Q'$ and $2Q$ phases and that between the $3Q$ and $2Q$ phases, 
indicating that less entropy is released at these transitions.
A small but nonzero negative value of $\chi$ is found only in the $3Q$ phase, as shown in Fig.~\ref{fig06}(b).
This indicates that when itinerant electrons are coupled with the $3Q$ spin texture, the system shows the topological Hall effect~\cite{Nagaosa2010}.

For the other field directions, ${\bf B}\parallel [110]$ and ${\bf B}\parallel [111]$,
the transition between the $2Q$ and $3Q$ phases in the low field regime takes place in a similar manner to that for ${\bf B}\parallel [100]$. 
In the $2Q$ phases,
$|\tilde{\bf M}_2| > |\tilde{\bf M}_1| >0$
for ${\bf B}\parallel [110]$,
while 
$|\tilde{\bf M}_3| > |\tilde{\bf M}_2| >0$,
$|\tilde{\bf M}_2| > |\tilde{\bf M}_1| >0$,
or $|\tilde{\bf M}_1| > |\tilde{\bf M}_3| >0$
for ${\bf B}\parallel [111]$.
For ${\bf B}\parallel [111]$, unlike the other magnetic field directions, 
the system undergoes the direct transition from the $3Q$ phase to the paramagnet, 
where $C$ drops suddenly similar to the phase transition between the $2Q$ and paramagnetic phases in Fig.~\ref{fig06}(b), 
while $\chi$ gradually vanishes similar to that between the $3Q$ and $2Q$ phases in Fig.~\ref{fig06}(b).
The order parameters are summarized in Table~\ref{tab1}.


Next, let us discuss the results for $(D,\Delta)=(0.15,0.3)$ in Fig.~\ref{fig05}, where the ground state at zero magnetic field is in the $3Q$ phase.
Unlike the previous case in Fig.~\ref{fig04}, 
there is only a second-order phase transition from the $3Q$ phase to the paramagnet at zero magnetic field while increasing temperature.
When we apply the magnetic field, although the system shows an overall common phase sequence from $3Q$ to $2Q$, and to $1Q$, 
there are qualitative differences depending on the magnetic field directions.

Figure~\ref{fig07} shows the field dependences of $|\tilde{\bf M}_\eta|$, 
$\chi$, and the magnetization per site, $m$, at $T=0.5$; 
$m$ is defined as
\begin{align}
m = \left| \frac{1}{\Lambda^3} \sum_{{\bf r}_0} \langle {\bf S}_{{\bf r}_0} \rangle \right|.
\end{align}
For ${\bf B}\parallel [100]$,
while increasing the magnetic field, we find successive transitions as $3Q \to 2Q \to 1Q \to$ paramagnet. 
All of them are of second order, 
where $|\tilde{\bf M}_3|$, $|\tilde{\bf M}_2|$, and $|\tilde{\bf M}_1|$ go to zero successively, as shown in Fig.~\ref{fig07}(a).
Nonzero $\chi$ is induced by the magnetic field in the $3Q$ phase; 
$\chi$ decreases almost linearly to $B$, 
but turns to increase around $B\simeq 0.626$ and 
vanishes at the transition between the $3Q$ and $2Q$ phases at $B\simeq 0.759$, 
as shown in Fig.~\ref{fig07}(b). 
The transition is accompanied by a kinklike anomaly in the magnetization curve.
In contrast, for ${\bf B}\parallel [110]$,  
we find two distinguishable double-$Q$ phases,
the $2Q$ and $2Q''$ phases, 
between the $3Q$ and $1Q$ phases. 
In the $2Q$ phase, $|\tilde{\bf M}_1|$ and $|\tilde{\bf M}_2|$ are nonzero, while in the $2Q''$ phase, $|\tilde{\bf M}_2|$ and $|\tilde{\bf M}_3|$ are nonzero, as shown in Fig.~\ref{fig07}(c). 
Notably, $\chi$ is zero for the former but becomes nonzero for the latter, as shown in Fig.~\ref{fig07}(d). 
The transitions are of second order for all the cases, except for that between the $2Q$ and $2Q''$ phases.
Finally, for ${\bf B}\parallel [111]$, the phase sequence is similar to that for ${\bf B}\parallel [100]$, 
but the transition between the $3Q$ and $2Q$ phases and that between the $2Q$ and $1Q$ phases are of both first order, as shown in Fig.~\ref{fig07}(e).
Note that $|\tilde{\bf M}_\eta|$ are exchangeable because of the $\mathsf{C}_3$ rotational symmetry about the [111] axis.
In this case, $\chi$ becomes nonzero not only in the $3Q$ phase but also in the $2Q$ phase, as shown in Fig.~\ref{fig07}(f).
We also summarize the order parameters in each phase found in these cases in Table~\ref{tab1}.

\subsection{$6Q$ model}\label{subsec:6Q}
Next, we discuss the model with six ${\bf Q}_\eta$, the $6Q$ model. 
We present the results in parallel with Sec.~\ref{subsec:3Q} for the $3Q$ model: 
After introducing the model parameters in Sec.~\ref{sec:modelparam6Q}, 
we present the ground-state phase diagram at zero magnetic field 
while varying $\Delta$ and $D$ in Sec.~\ref{sec:GroundState6Q}, and then, 
the magnetic field-temperature phase diagrams for a couple of representative parameter sets of $\Delta$ and $D$ in Sec.~\ref{sec:FiniteT6Q}.

\subsubsection{Model parameters}\label{sec:modelparam6Q}
\begin{figure}[bh]
  \centering
  \includegraphics[trim=0 0 0 0, clip,width=\columnwidth]{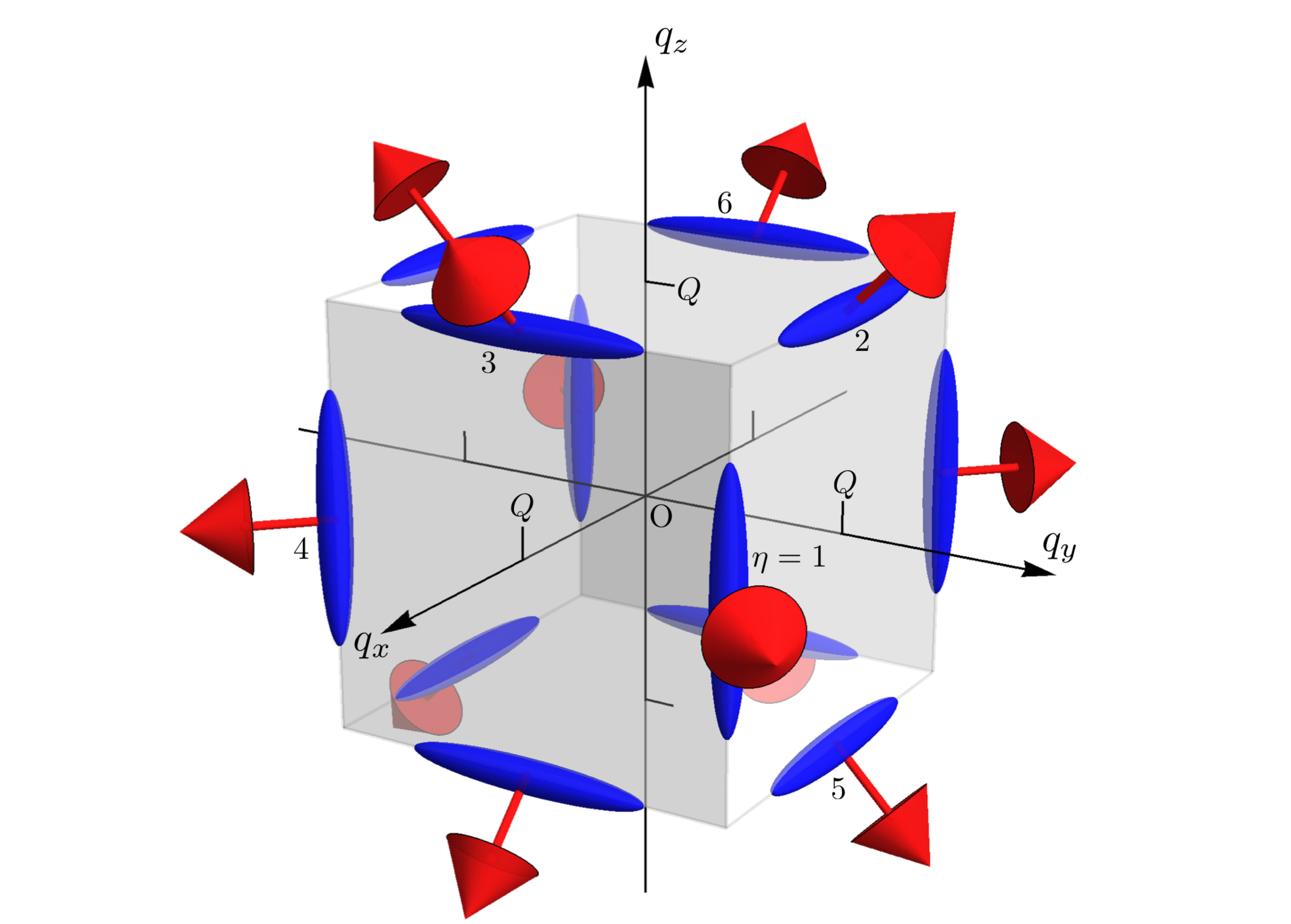}
  \caption{
  Pictorial representation of the coupling constants for the symmetric and antisymmetric interactions in the $6Q$ model.
  The notations are common to those in Fig.~\ref{fig01}. 
  }
  \label{fig08}
\end{figure}
The model Hamiltonian is given by Eqs.~\eqref{eq:H} and \eqref{eq:Hq} with $N_Q=6$ with the set of ${\bf Q}_\eta$:
\begin{align}
&{\bf Q}_1 = Q (\hat{\bf x} + \hat{\bf y}),~
{\bf Q}_2 = Q (\hat{\bf y} + \hat{\bf z}),~
{\bf Q}_3 = Q(\hat{\bf z} + \hat{\bf x}), \nonumber \\
&{\bf Q}_4 = Q(\hat{\bf x} - \hat{\bf y}),~
{\bf Q}_5 = Q(\hat{\bf y} - \hat{\bf z}),~
{\bf Q}_6 = Q(\hat{\bf z} - \hat{\bf x}).~\label{eq:Q6Q}
\end{align}
Similar to the $3Q$ model, we take $Q=2\pi/\Lambda$ with $\Lambda = 12$. 
We set $J_\eta^{\alpha \alpha}$ in Eq.~\eqref{eq:JD} as
\begin{align}
&(J_\eta^{xx},J_\eta^{yy},J_\eta^{zz}) = \nonumber\\
&\begin{cases}
 [J(1-\Delta), J (1-\Delta), J(1+2\Delta)],&(\eta=1,4)\\
 [J(1+2\Delta), J(1-\Delta), J (1-\Delta)],&(\eta=2,5)\\
 [J (1-\Delta), J(1+2\Delta), J(1-\Delta)],&(\eta=3,6)\\
 \end{cases}. 
\label{eq:J6Q}
\end{align}
Figure~\ref{fig08} shows the pictorial representation of $J_\eta^{\alpha \alpha}$ and ${\bf D}_\eta$.
We take the energy unit as $J=1$ as before.

\subsubsection{Ground state at zero magnetic field}\label{sec:GroundState6Q}
\begin{figure}[htbp]
  \centering
  \includegraphics[trim=0 0 0 0, clip,width=\columnwidth]{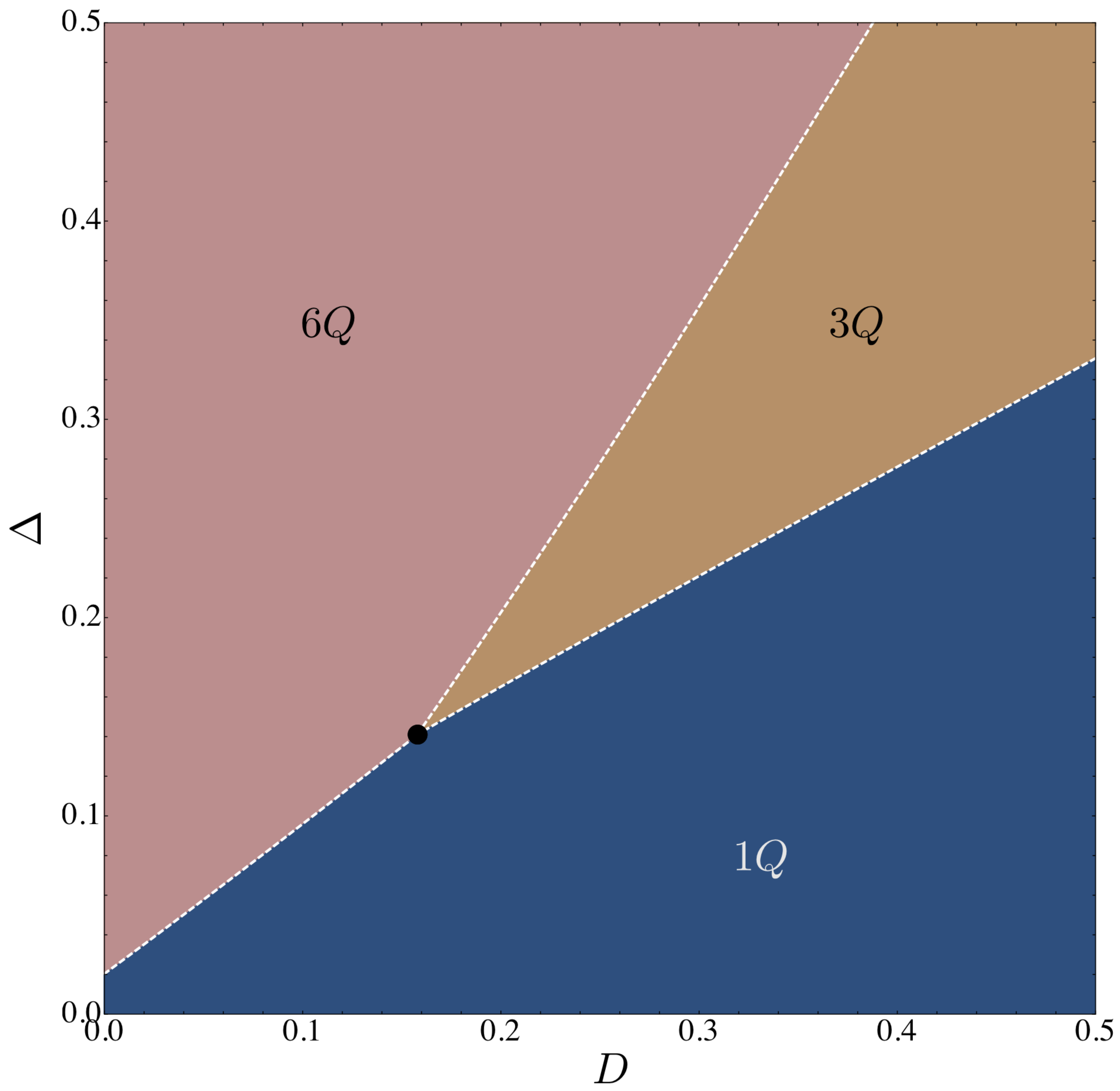}
  \caption{
  Ground-state phase diagram for the $6Q$ model at zero magnetic field.
  The phase diagram includes the $1Q$ phase where one of $|\tilde{\bf M}_{\eta}|$ is nonzero, 
  the $3Q$ phase where three of $|\tilde{\bf M}_\eta|$ are nonzero,
  and the $6Q$ phase where all six $|\tilde{\bf M}_\eta|$ are nonzero. 
  In the $3Q$ and $6Q$ phases, all the nonzero $|\tilde{\bf M}_\eta|$ take the same value, while the value depends on $D$ and $\Delta$. 
  The white dashed lines separating these phases represent the first-order phase transitions, 
  and the black dot locates the triple point where the three first-order transition lines meet.
  }
  \label{fig09}
\end{figure}
\begin{figure}[htbp]
  \centering
  \includegraphics[trim=0 0 0 0, clip,width=0.95 \columnwidth]{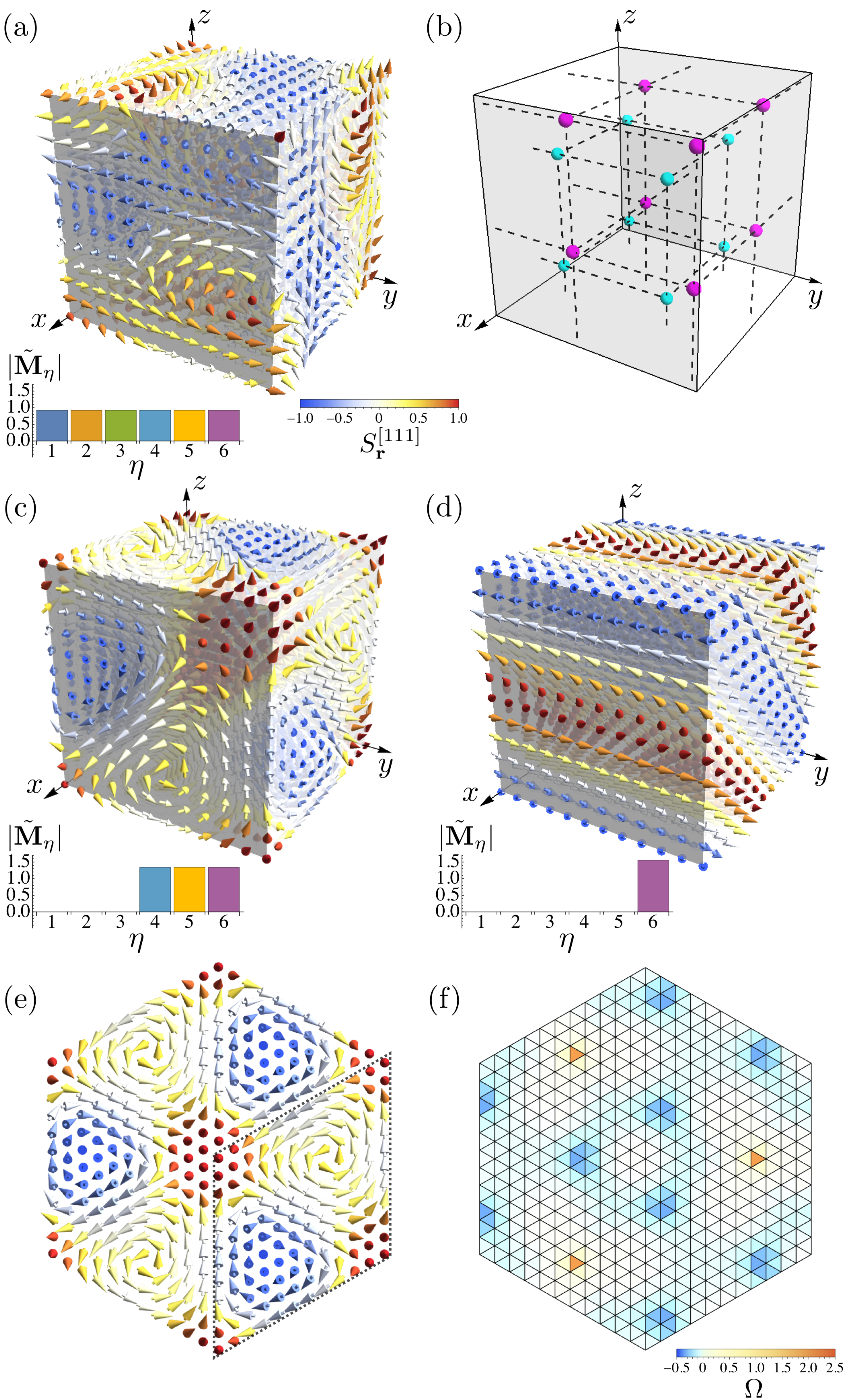}
  \caption{
  Ground-state spin configurations stabilized in the $6Q$ model at zero magnetic field for 
   (a) the $6Q$ state at $(D,\Delta)=(0.2,0.3)$,
   (c) the $3Q$ state at $(D,\Delta)=(0.3,0.3)$, and
   (d) the $1Q$ state at $(D,\Delta)=(0.1,0.0)$.
   The color of the arrows denotes the $[111]$ component of spins, 
   $S^{[111]}_{\bf r}$, according to the color bar in (a). 
   (b) Positions of the hedgehogs (magenta spheres) and the antihedgehogs (cyan spheres) in the $6Q$ state shown in (a). 
   The dashed lines are guides to the eye.
   Insets of (a), (c), and (d) show distributions of $|\tilde{\bf M}_\eta|$ for each state. 
   (e) Spin configuration on a $(111)$ slice of (c). 
   The black dashed rhombus indicates the 2D magnetic unit cell.
   (f) Distribution of the solid angle $\Omega$ spanned by neighboring three spins calculated from the spin configuration in (e).
  }
  \label{fig10}
\end{figure}
\begin{figure*}[htbp]
  \centering
  \includegraphics[trim=0 0 0 0, clip,width=\textwidth]{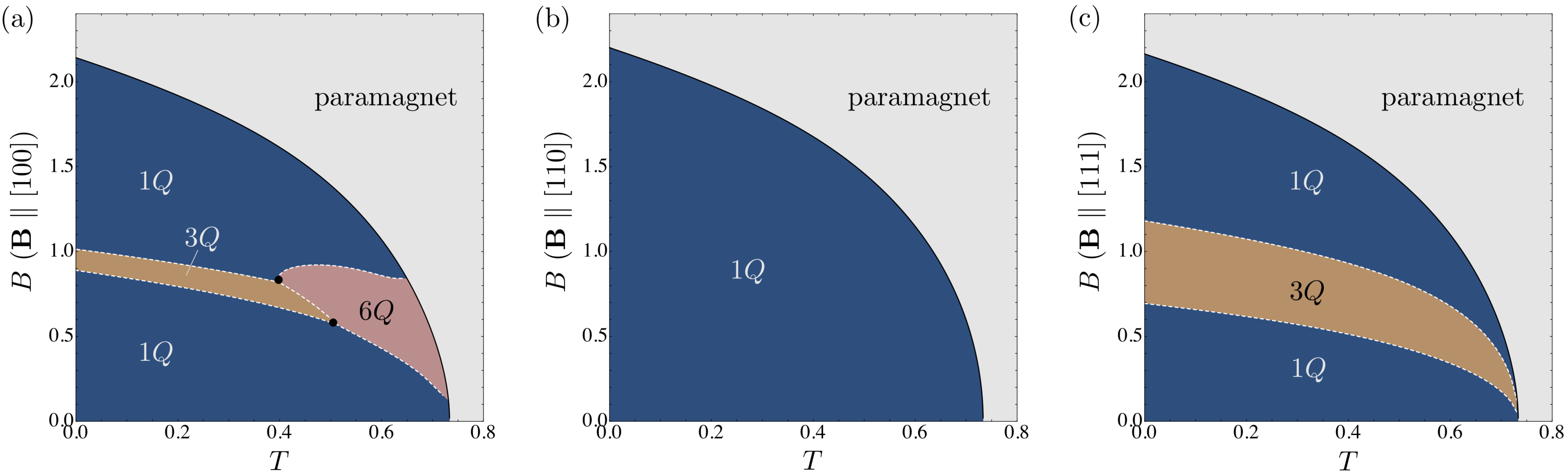}
  \caption{
  Magnetic field-temperature phase diagrams of the $6Q$ model with $(D,\Delta)=(0.1,0.0)$ for the magnetic field directions
  (a) ${\bf B}\parallel [100]$,
  (b) ${\bf B}\parallel [110]$, and
  (c) ${\bf B}\parallel [111]$.
  The white dashed lines and the black solid lines represent first-order and second-order phase transitions, respectively.
  The black dots in (a) locate the triple points.
  }
  \label{fig11}
\end{figure*}
\begin{figure*}[htpb]
  \centering
  \includegraphics[trim=0 0 0 0, clip,width=\textwidth]{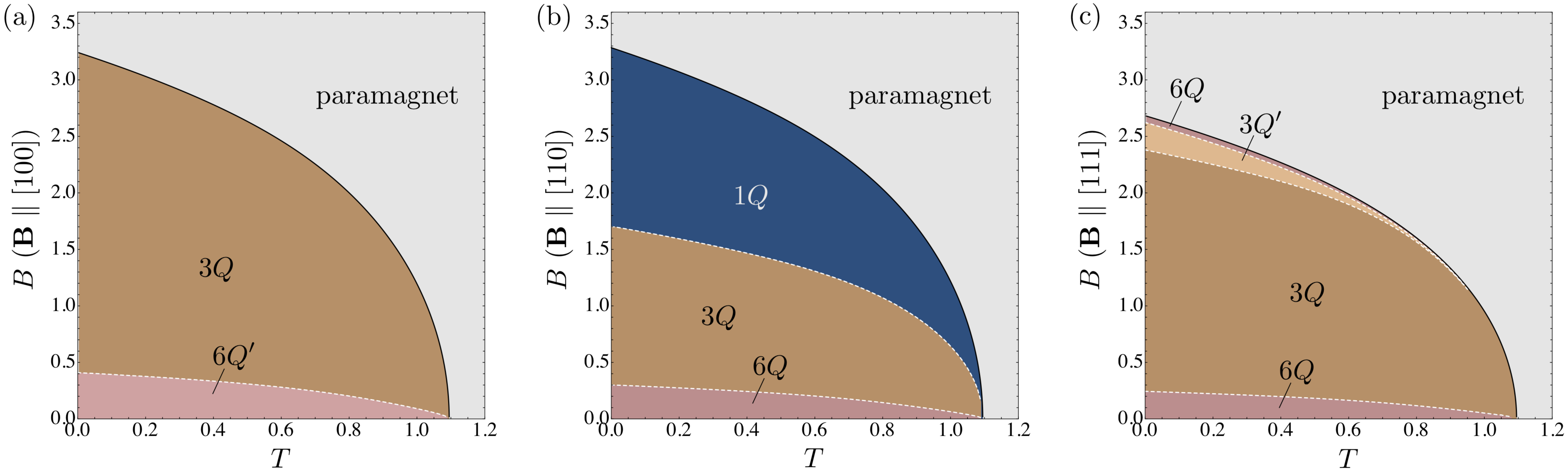}
  \caption{
  Magnetic field-temperature phase diagrams of the $6Q$ model with $(D,\Delta)=(0.2,0.3)$ for the magnetic field directions
  (a) ${\bf B}\parallel [100]$,
  (b) ${\bf B}\parallel [110]$, and
  (c) ${\bf B}\parallel [111]$.
  The notations are common to those in Fig.~\ref{fig11}.
  }
  \label{fig12}
\end{figure*}
\begin{figure}[htpb]
  \centering
  \includegraphics[trim=0 0 0 0, clip,width=\columnwidth]{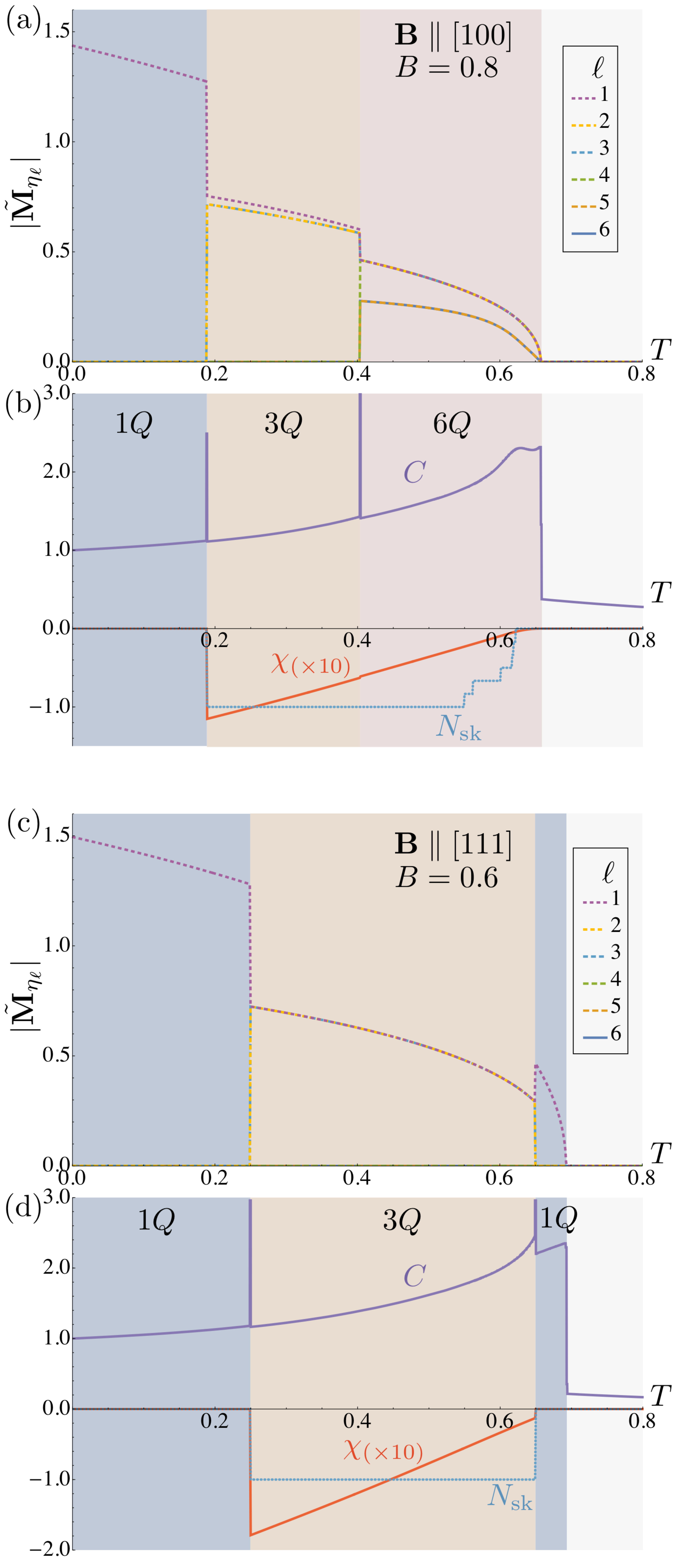}
  \caption{
  Temperature dependences of (a,c) the order parameters $|\tilde{\bf M}_{\eta_\ell}|$ and 
  (b,d) the specific heat $C$, the spin scalar chirality $\chi$, and the skyrmion number $N_{\rm sk}$ for the $6Q$ model with $(D,\Delta)=(0.1,0.0)$. 
  The data in (a,b) are for ${\bf B}\parallel [100]$ with $B=0.8$, 
  while those in (c,d) are for ${\bf B}\parallel [111]$ with $B=0.6$. 
   The order parameters are sorted in descending order:
   $|\tilde{\bf M}_{\eta_1}| \geq |\tilde{\bf M}_{\eta_2}| \geq |\tilde{\bf M}_{\eta_3}| \geq |\tilde{\bf M}_{\eta_4}| \geq |\tilde{\bf M}_{\eta_5}| \geq |\tilde{\bf M}_{\eta_6}| $.
   The spin scalar chirality is multiplied by a factor of $10$ for better visibility.
  }
  \label{fig13}
\end{figure}
\begin{figure*}[htpb]
  \centering
  \includegraphics[trim=0 0 0 0, clip,width=\textwidth]{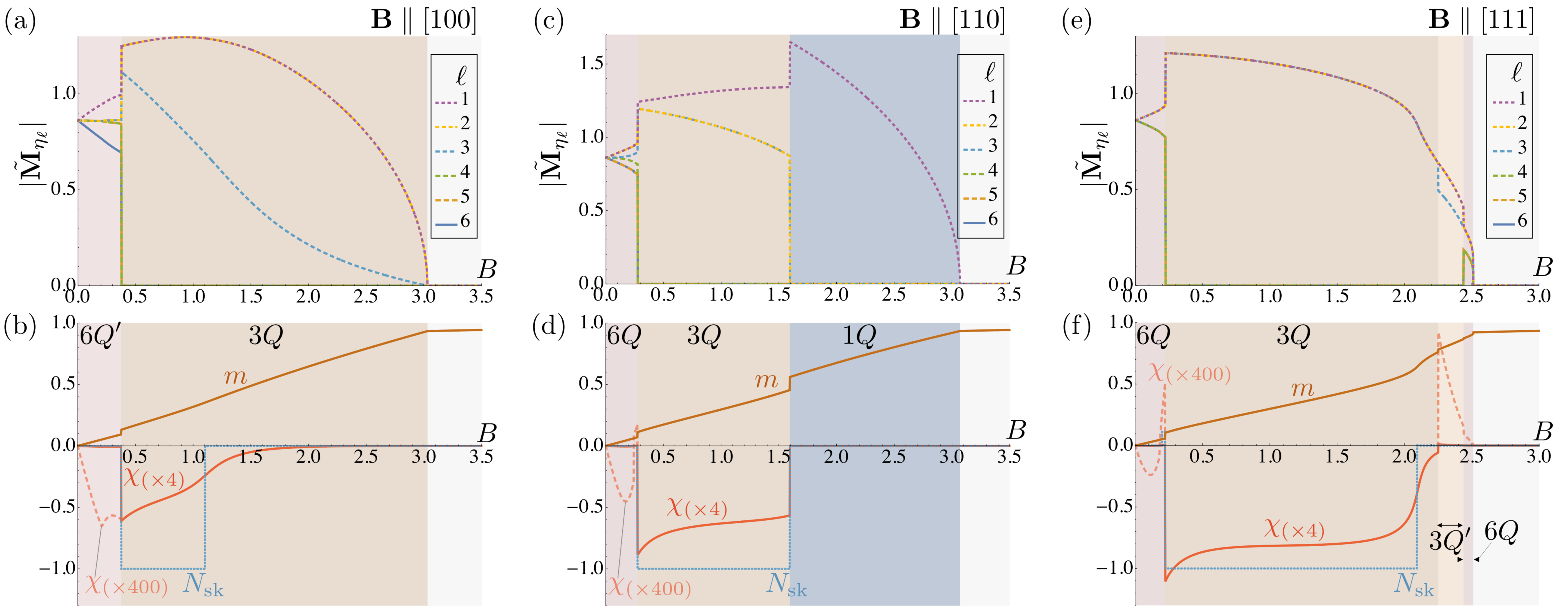}
  \caption{
  Magnetic field dependences of (a,c,d) the order parameters $|\tilde{\bf M}_{\eta_\ell}|$ and 
  (b,d,f) the magnetization $m$, 
  the spin scalar chirality $\chi$, and the skyrmion number $N_{\rm sk}$ for the $6Q$ model with $(D,\Delta)=(0.2,0.3)$ and 
  $T=0.2$. 
   The magnetic field directions are (a,b) ${\bf B}\parallel [100]$, 
   (c,d) ${\bf B}\parallel [110]$, and (e,f) ${\bf B}\parallel [111]$. 
   The order parameters are sorted in descending order as Fig.~\ref{fig12}.
   The spin scalar chiralities in (b), (d), and (f) are multiplied by a factor of $4$ (solid lines)
   [a factor of $400$ in the $6Q$ and $6Q'$ phases (dashed lines)] for better visibility.
  }
  \label{fig14}
\end{figure*}
\begin{figure}[htpb]
  \centering
  \includegraphics[trim=0 0 0 0, clip,width=\columnwidth]{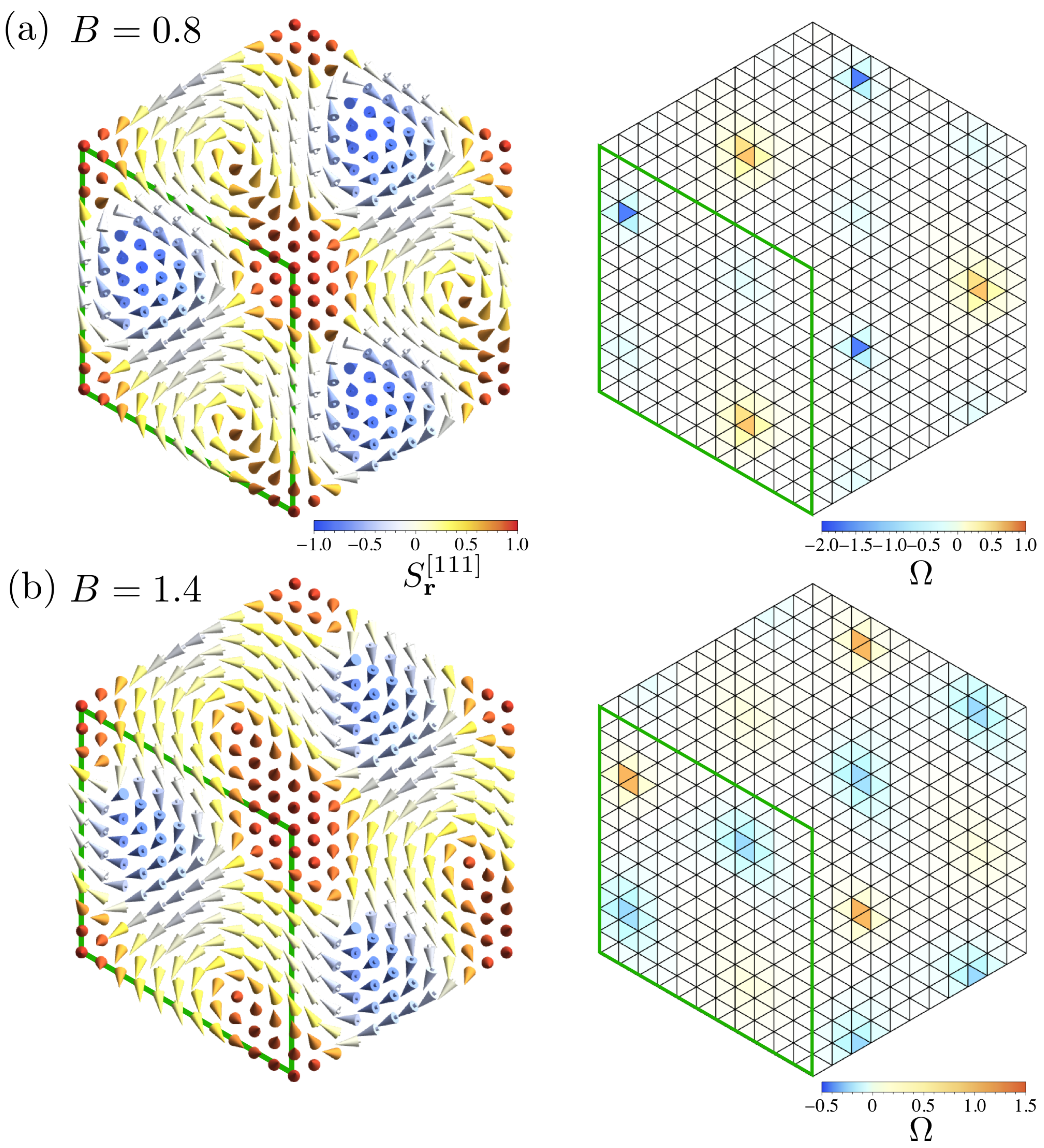}
  \caption{
  Spin configurations (left) and distribution of the solid angle $\Omega$ (right) in the $6Q$ model with $(D,\Delta)=(0.2,0.3)$ and 
  $T=0.2$ in ${\bf B}\parallel [100]$: (a) $B=0.8$ and (b) $B=1.4$, which correspond to before and after the topological transition in the $3Q$ phase, respectively. 
  The notations are common to those in Figs.~\ref{fig10}(e) and \ref{fig10}(f).  
  The green rhombi correspond to a $(100)$ slice of the magnetic unit cell.
  }
  \label{fig15}
\end{figure}
\begin{figure}[htpb]
  \centering
  \includegraphics[trim=0 0 0 0, clip,width=\columnwidth]{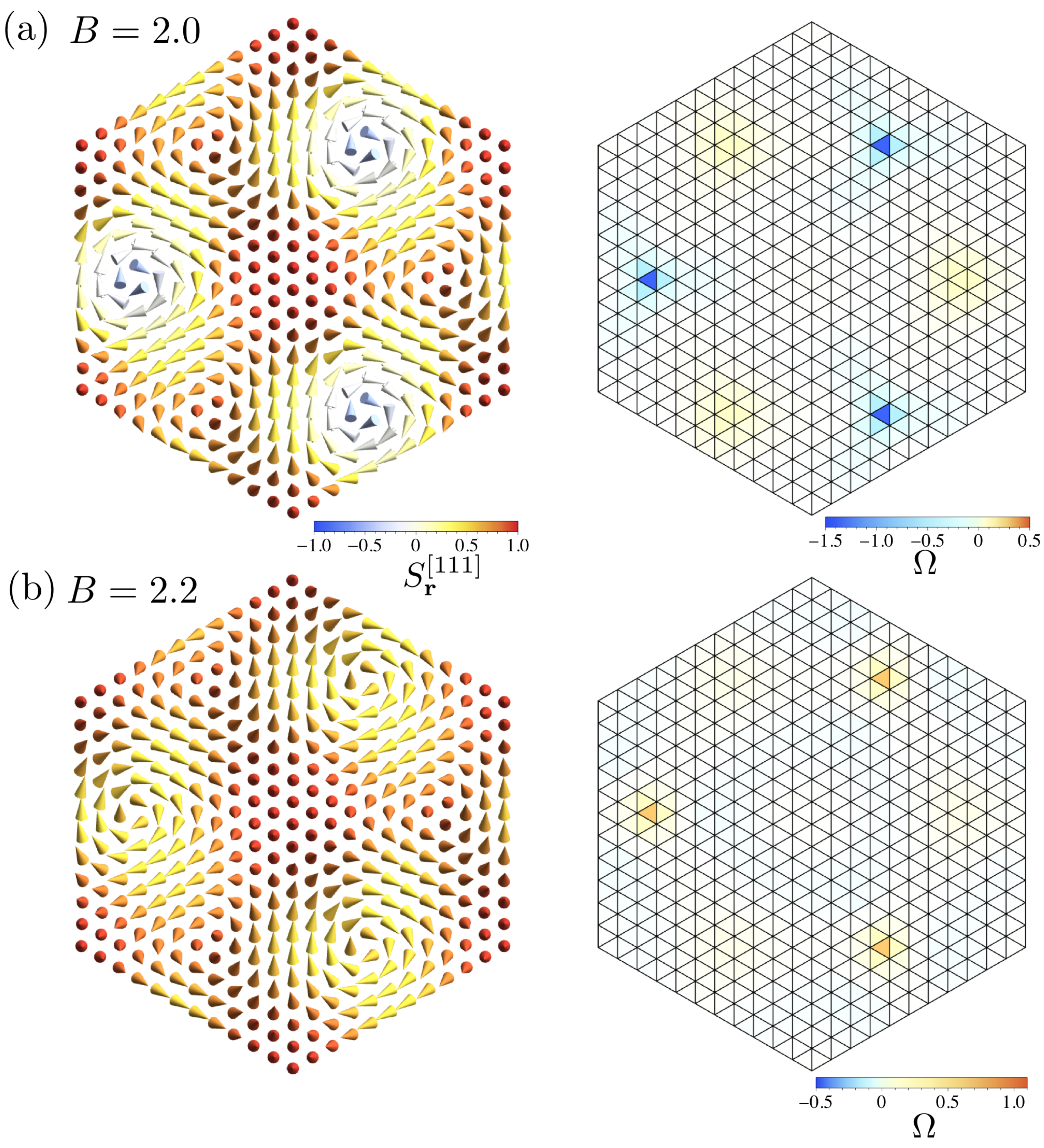}
  \caption{
  Spin configurations (left) and distribution of the solid angle $\Omega$ (right) in the $6Q$ model with $(D,\Delta)=(0.2,0.3)$ and 
  $T=0.2$ in ${\bf B}\parallel [111]$: (a) $B=2.0$ and (b) $B=2.2$, which correspond to before and after the topological transition in the $3Q$ phase, respectively. 
  The notations are common to those in Fig.~\ref{fig15}. 
  }
  \label{fig16}
\end{figure}

Figure~\ref{fig09} shows the ground-state phase diagram for the $6Q$ model at zero magnetic field while changing $\Delta$ and $D$.
We find three stable phases, similarly to the $3Q$ model~(see Fig.~\ref{fig02}): 
The $1Q$ phase in the small $\Delta$ region, the $6Q$ phase in the large $\Delta$ region,
and the $3Q$ phase in between them.
Unlike the $3Q$ model, however, all the transitions are discontinuous, and furthermore, 
the intermediate $3Q$ phase does not extend down to $D=0$. 
This results in the triple point denoted by the black dot in Fig.~\ref{fig09}, 
where the three first-order transition lines meet.

We display typical spin textures in the three phases in Fig.~\ref{fig10},
with the values of $|\tilde{\bf M}_\eta|$ in each inset.
First, Fig.~\ref{fig10}(a) represents the $6Q$ state.
This is a 3D HL, in which the magnetic hedgehogs and antihedgehogs forming a periodic lattice, as shown in Fig.~\ref{fig10}(b).
In this phase, similar to the $3Q$ state in the $3Q$ model~(see Sec.~\ref{sec:GroundState3Q}), 
$|\tilde{\bf M}_\eta| $ for all $\eta$ are the same;
namely, the $6Q$ state is composed of a superposition of six proper screws with equal amplitudes. 
In this phase, the $\mathsf{C}_3$ rotational symmetry about the $\langle111\rangle$ axis is retained.
We note that the $6Q$ state is closely related to the magnetic state called bcc2 in a phenomenological Ginzburg-Landau theory in Refs.~\cite{Binz2006a,Binz2006b,Binz2008}.
This is explicitly confirmed by calculating
$\tilde{T}_x = -\tilde{M}^z_1 (\tilde{M}^y_3)^* \tilde{M}^z_4 \tilde{M}^y_6  $, 
$\tilde{T}_y= -(\tilde{M}^z_1)^* \tilde{M}^x_2 \tilde{M}^z_4 \tilde{M}^x_5  $, and 
$\tilde{T}_z= -(\tilde{M}^x_2)^* \tilde{M}^y_3 \tilde{M}^x_5 \tilde{M}^y_6  $, 
corresponding to $T_x$, $T_y$, and $T_z$, respectively, discussed in the previous studies; 
we confirm that all $\tilde{T}_\mu $ are positive real as $T_\mu$ in the bcc2 state.
A difference is that the superposed spin helices of the $6Q$ state are elliptically distorted due to the anisotropy, 
whereas those of the bcc2 state are not distorted.

Next, Fig.~\ref{fig10}(c) represents the $3Q$ state.
This state is composed of a superposition of three proper screws with equal amplitudes, 
in which the possible combinations of $\eta$ for nonzero $|\tilde{\bf M}_\eta|$ are limited to 
$\{\eta_1, \eta_2, \eta_3\} $ satisfying 
${\bf Q}_{\eta_1} \pm {\bf Q}_{\eta_2} \pm {\bf Q}_{\eta_3} =0$, 
namely, $\{\eta_1, \eta_2, \eta_3\} = \{1,2,6\}$, $\{1,3,5\}$, $\{2,3,4\}$, or $\{4,5,6\}$. 
For example, in the case of $\{\eta_1, \eta_2, \eta_3\} = \{4,5,6\}$, for which ${\bf Q}_4+{\bf Q}_5+{\bf Q}_6=0$,
as all the three ${\bf Q}_{\eta}$ are orthogonal to the $[111]$ axis, 
there is no spin modulation in the $[111]$ direction: 
Any $(111)$ slice gives the same spin configuration regardless of the position of the cut. 
Interestingly, the 2D spin texture on the $(111)$ slice is topologically nontrivial. 
Figures~\ref{fig10}(e) and \ref{fig10}(f) show the spins configuration and
the distribution of the corresponding solid angle formed by neighboring three spins, $\Omega$, respectively (refer to Ref.~\cite{Okumura2020} for the calculation of $\Omega$).
The results indicate that this $3Q$ state is a SkX with skyrmion number  
$N_{\rm sk} = -1$. 
This is explicitly shown by summing up $\Omega$ in Fig.~\ref{fig10}(f) within the 2D magnetic unit cell denoted by the dashed rhombus in Fig.~\ref{fig10}(e). 
Thus, the $3Q$ state consists of 2D SkXs stacked along the $[111]$ direction. 
Note that this state is energetically degenerate with the stacking of SkXs with $N_{\rm sk}=+1$ obtained by flipping all the spins.
Such a stacked topological spin structure is common to other combinations of $\{\eta_1, \eta_2, \eta_3\}$ listed above, 
while a particular set (or subset) with a particular value of $N_{\rm sk}$ might
be energetically favored when a magnetic field is applied.
We note that the $\mathsf{C}_3$ rotational symmetry about the $\langle 111\rangle$ axis is weakly broken in this $3Q$ phase even at zero magnetic field.

Lastly, Fig.~\ref{fig10}(d) represents the $1Q$ state where only one of $|\tilde{\bf M}_\eta|$ is nonzero. 
The nonzero component of $|\tilde{\bf M}_\eta|$ can be chosen arbitrarily among the six at zero magnetic field, while a particular one (or one from a particular subset) will be selected in an applied magnetic field depending on its direction.

\subsubsection{Magnetic field-temperature phase diagrams}\label{sec:FiniteT6Q}
Figures~\ref{fig11} and \ref{fig12} show the magnetic field-temperature phase diagrams for the representative parameter sets
that realize the $1Q$ and $6Q$ ground states at zero magnetic field, respectively.
We take $(D,\Delta)=(0.1,0.0)$ for the $1Q$ case and $(D,\Delta)=(0.2,0.3)$ for the $6Q$ case for which 
the ground-state spin configurations are shown in Figs.~\ref{fig10}(d) and \ref{fig10}(a), respectively.
Similar to the analysis of the $3Q$ model in Sec.~\ref{sec:FiniteT3Q},
in each case, we obtain the results for different magnetic field directions, 
${\bf B}\parallel [100]$, ${\bf B}\parallel [110]$, and ${\bf B}\parallel [111]$ in panels (a), (b), and (c), respectively, of Figs.~\ref{fig11} and \ref{fig12}.

Let us begin with the results for $(D,\Delta)=(0.1,0.0)$ in Fig.~\ref{fig11}. 
At zero magnetic field, the system is in the $1Q$ phase below the critical temperature at $T \simeq 0.733$; the phase transition between the $1Q$ and paramagnetic phases is of second order. 
In an applied magnetic field, the phase diagram is qualitatively different depending on the direction of the magnetic field. 
While there is no additional phase for ${\bf B}\parallel [110]$ [Fig.~\ref{fig11}(b)], 
we find phase transitions to multiple-$Q$ phases for ${\bf B}\parallel [100]$ [Fig.~\ref{fig11}(a)] and ${\bf B}\parallel [111]$ [Fig.~\ref{fig11}(c)]. 
In the case of ${\bf B}\parallel [100]$, the system changes from the $1Q$ phase to the $3Q$ 
and $6Q$ phases in the intermediate magnetic field region at low and high temperature,
respectively, and comes back to the $1Q$ phase for a higher magnetic field; 
namely, the system undergoes reentrant transitions between the single- and multiple-$Q$ phases.
All the transitions between the magnetically ordered phases are discontinuous, 
resulting in the two triple points denoted by the black dots in Fig.~\ref{fig11}(a). 
Notably, the $6Q$ phase appears only at finite temperature, and the width  becomes wider for higher temperature, suggesting that it is stabilized by the entropic gain, similar to the $3Q$ phase in the $3Q$ model in Fig.~\ref{fig04}. 
In contrast, for ${\bf B}\parallel [111]$, we find a reentrant transition as $1Q\to 3Q\to 1Q$, as shown in Fig.~\ref{fig11}(c), where the intermediate $3Q$ phase becomes narrower while increasing temperature and vanishes into the transition point between the $1Q$ and paramagnetic phases in the zero-field limit.
It is worth noting that these results are for $\Delta=0$: 
The multiple-$Q$ phases are stabilized under the magnetic field even in the absence of the magnetic anisotropy in the symmetric exchange interactions.

Figures~\ref{fig13}(a) and \ref{fig13}(b) show the temperature dependences of 
$|\tilde{\bf M}_\eta|$, $C$, $\chi$, and $N_{\rm sk}$ in the intermediate magnetic-field regime for ${\bf B}\parallel [100]$ ($B=0.8$). 
In this case, the system undergoes successive transitions as $1Q \to 3Q \to 6Q \to$ paramagnet while increasing temperature. 
In the $1Q$ phase at low temperature, $\eta$ for the nonzero $|\tilde{\bf M}_{\eta}|$ is chosen from $1$, $3$, $4$, or $6$
for which the easy axis in the corresponding symmetric interaction is perpendicular to ${\bf B}$ (see Fig.~\ref{fig08}). 
Meanwhile, in the intermediate $3Q$ phase, 
three out of six $|\tilde{\bf M}_{\eta}|$ are nonzero 
with the relation $|\tilde{\bf M}_{\eta_1}| > |\tilde{\bf M}_{\eta_2}| =|\tilde{\bf M}_{\eta_3}| > 0 $ 
where $\{ \eta_1, \eta_2,\eta_3\}$ are chosen from $\{ 2, 1, 6\}$, $\{5, 1,3\}$, $\{2, 3,4\}$, or $\{5,4,6\}$. 
In the $6Q$ phase at high temperature,  all of $|\tilde{\bf M}_{\eta}|$ are nonzero with 
the relation $|\tilde{\bf M}_1| = |\tilde{\bf M}_3| = |\tilde{\bf M}_4| = |\tilde{\bf M}_6| \geq |\tilde{\bf M}_{2}|  = |\tilde{\bf M}_{5}| > 0$.
We summarize the order parameters in each phase in Table~\ref{tab2}. 
The two transitions between the magnetically ordered phases are both of first order, as indicated by the delta-function type anomalies in $C$ shown in Fig.~\ref{fig13}(b). 
Meanwhile, the transition from the $6Q$ phase to the paramagnet is continuous, where $C$ shows a jump, similar to the case of the $3Q$ model in Fig.~\ref{fig06}(b).

We note that, at the phase transition from the $6Q$ phase to the paramagnet,
the six components of the order parameters $|\tilde{\bf M}_\eta|$ show different critical behaviors: 
Four out of them go to zero in a square root fashion, but the rest two vanish
linearly, as shown in Fig.~\ref{fig13}(a).
These peculiar behaviors are understood from the expansion of $g (\{ \overline{ \tilde{m}_{{\bf r}_0}^\alpha } \})$ in Eq.~\eqref{eq:g} in terms of $\tilde{M}^\alpha_\eta$ in Eq.~\eqref{eq:m2M}, 
which corresponds to the Ginzburg--Landau theory.
Among the relevant contributions to the stabilization of the $6Q$ phase, we obtain a 
third-order term given by
\begin{align}
&B \bigr[ 
\tilde{M}^x_{2} \{
(\tilde{\bf M}_1 \cdot \tilde{\bf M}_6)^*
+
(\tilde{\bf M}_3^* \cdot \tilde{\bf M}_4)
\} \nonumber\\
&+
\tilde{M}^x_{5} \{
(\tilde{\bf M}^*_1 \cdot \tilde{\bf M}_3) \:\:
+
(\tilde{\bf M}_4 \cdot \tilde{\bf M}_6)
\}
+{\rm c.c.} \bigr],
\label{eq:GL_6Q}
\end{align}
which represents the coupling among ${\bf B}$, 
the $x$ component of $\tilde{\bf M}_\eta$ with $\eta = 2$ or $5$, 
and the other two $\tilde{\bf M}_\eta$ with $\eta', \eta'' = 1$, $3$, $4$, or $6$ that satisfy ${\bf Q}_\eta \pm {\bf Q}_{\eta'} \pm {\bf Q}_{\eta''} =0$~(see Fig.~\ref{fig08}).
Given this form, our result in Fig.~\ref{fig13}(a) indicates that $\tilde{\bf M}_{1}$, $\tilde{\bf M}_{3}$, $\tilde{\bf M}_{4}$, and  $\tilde{\bf M}_{6}$ are the primary order parameters, 
and $\tilde{\bf M}_{2}$ and $\tilde{\bf M}_{5}$ are the secondary ones:
\begin{align}
&|\tilde{\bf M}_1| = |\tilde{\bf M}_3|  = |\tilde{\bf M}_4|  = |\tilde{\bf M}_6| \propto |T-T_c|^{1/2},\\
&|\tilde{\bf M}_2| = |\tilde{\bf M}_5| \propto |T-T_c|,
\end{align}
near the critical temperature $T_c$.
Thus, 
$B \{ (\tilde{\bf M}_1 \cdot \tilde{\bf M}_6)^*+(\tilde{\bf M}_3^* \cdot \tilde{\bf M}_4) \}$
and 
$B \{ (\tilde{\bf M}^*_1 \cdot \tilde{\bf M}_3) + (\tilde{\bf M}_4 \cdot \tilde{\bf M}_6) \}$, 
both of which are proportional to $|T-T_c|$, 
act as internal fields to induce $\tilde{M}^x_{2}$ and $\tilde{M}^x_{5}$, respectively, through Eq.~\eqref{eq:GL_6Q}. 
At the same time, this analysis indicates that a nonzero magnetic field plays a key role for the stabilization of the $6Q$ phase in Fig.~\ref{fig11}(a), in contrast to the $3Q$ phases in Fig.~\ref{fig04}.

As shown in Fig.~\ref{fig13}(b), $\chi$ becomes nonzero in the $3Q$ and $6Q$ phases. 
Notably, the absolute value is almost two or three orders of magnitude larger than that in the $3Q$ phase in the $3Q$ model [see Figs.~\ref{fig06}(b), \ref{fig07}(b), \ref{fig07}(d), and \ref{fig07}(f)]. 
This is because the $3Q$ state in Figs.~\ref{fig13}(a) and \ref{fig13}(b) is topologically nontrivial, which consists of stacked SkXs with $N_{\rm sk}=-1$, similar to the state at zero magnetic field in Figs.~\ref{fig10}(e) and \ref{fig10}(f). 
Note that the zero field state is energetically degenerate between $N_{\rm sk}=+1$ and $-1$, but the one with $N_{\rm sk}=-1$ is energetically preferred under the magnetic field. 
As shown in Fig.~\ref{fig13}(b), $N_{\rm sk}$ remains at $-1$ in the high-$T$ $6Q$ phase, but it increases in a stepwise manner, according to the motions of hedgehogs and antihedgehogs on the discrete lattice. 
Note that $N_{\rm sk}$ is an average over the $(100)$ slices, some of which has $N_{\rm sk}=-1$ and the others have $N_{\rm sk}=0$ depending on how many Dirac strings connecting the hedgehogs and antihedgehogs penetrate the slice. 
Finally, $N_{\rm sk}$ goes to zero at $T \simeq 0.622$, where the hedgehogs and antihedgehogs cause pair annihilation. 
This is a topological transition caused by temperature, whose remnant can be seen as a hump in the specific heat in Fig.~\ref{fig13}(a). 

\begin{table}[hbtp]
\caption{\label{tab2}
Order parameters in each magnetically ordered phase of the $6Q$ model. 
The sets of $\eta$, $\{\eta_1,\eta_2,\cdots\}$, for nonzero $\tilde{\bf M}_\eta$ are shown, with the relation between $|\tilde{\bf M}_\eta|$.
}
\begin{ruledtabular}
\begin{tabular}{lccl}
                   & Phase  & Sets of $\eta$ & Notes \\
\hline
${\bf B}=0$ & $1Q$ & $\{1\}, \{2\}, \{3\}$, &\\
                  &            & $\{4\}, \{5\}, \{6\}$ &\\
		  & $3Q$ & $\{1,2,6\}, \{1,3,5\}$,  & $|\tilde{\bf M}_{\eta_1} | =|\tilde{\bf M}_{\eta_2} |=|\tilde{\bf M}_{\eta_3} |$\\ 
		  &           & $\{2,3,4\}, \{4,5,6\}$ & \\ 
                   & $6Q$ & $\{1,2,3,4,5,6\}$ & $|\tilde{\bf M}_{\eta_1} | = |\tilde{\bf M}_{\eta_2} |= |\tilde{\bf M}_{\eta_3} |$\\
                   &  &                           & $= |\tilde{\bf M}_{\eta_4} |= |\tilde{\bf M}_{\eta_5} |= |\tilde{\bf M}_{\eta_6} |$
 \vspace{0.1cm}\\
\hline
${\bf B}\parallel [100]$ & $1Q$ & $\{1\}, \{3\}, \{4\}, \{6\}$ &\\
                   & $3Q$ & $\{1,6,2\}, \{1,3,5\}$,  & $|\tilde{\bf M}_{\eta_1}| =|\tilde{\bf M}_{\eta_2}| \neq |\tilde{\bf M}_{\eta_3}| $ \\ 
                   &  & $\{3,4,2\}, \{4,6,5\}$ &  \\ 
                   & $6Q$ & $ \{1,3,4,6,2,5\}
$, & $|\tilde{\bf M}_{\eta_1}| = |\tilde{\bf M}_{\eta_2}|  = |\tilde{\bf M}_{\eta_3}|$\\
                   &   &  & $= |\tilde{\bf M}_{\eta_4}| > |\tilde{\bf M}_{\eta_5}|=|\tilde{\bf M}_{\eta_6}|$\\
                   & $6Q'$ & $\{2,4,6,1,3,5\}$, & $|\tilde{\bf M}_{\eta_1}| > |\tilde{\bf M}_{\eta_2}| =|\tilde{\bf M}_{\eta_3}|$\\
                   &   & $\{5,1,3,4,6,2\}$& $\geq |\tilde{\bf M}_{\eta_4}|=|\tilde{\bf M}_{\eta_5}|>|\tilde{\bf M}_{\eta_6}|$
 \vspace{0.1cm}\\
\hline
${\bf B}\parallel [110]$ & $1Q$  & $\{1\}$ &\\
                   & $3Q$ & $\{4,5,6\}, \{4,2,3\}$  & $|\tilde{\bf M}_{\eta_1} | > |\tilde{\bf M}_{\eta_2} |=|\tilde{\bf M}_{\eta_3} |$\\ 
                   & $6Q$ & $\{2,3,4,1,5,6\}$, & $|\tilde{\bf M}_{\eta_1}| = |\tilde{\bf M}_{\eta_2}| > |\tilde{\bf M}_{\eta_3}| $\\
                   &          & $\{5,6,4,1,2,3\}$ & $ > |\tilde{\bf M}_{\eta_4}|>|\tilde{\bf M}_{\eta_5}|=|\tilde{\bf M}_{\eta_6}|$
 \vspace{0.1cm}\\
\hline
${\bf B}\parallel [111]$ & $1Q$ & $\{1\}$,$\{2\}$,$\{3\}$ &\\
                   & $3Q$ & $\{4,5,6\}$ & $|\tilde{\bf M}_{\eta_1} | = |\tilde{\bf M}_{\eta_2} |= |\tilde{\bf M}_{\eta_3} |$\\ 
                   & $3Q'$ &  $\{1,2,6\}, \{2,3,4\},$ & $|\tilde{\bf M}_{\eta_1} | = |\tilde{\bf M}_{\eta_2} | >  |\tilde{\bf M}_{\eta_3} | $\\ 
                   &  &  $\{1,3,5\}$ & \\ 
                   & $6Q$ & $\{4,5,6,1,2,3\}$ & $|\tilde{\bf M}_{\eta_1} | = |\tilde{\bf M}_{\eta_2} |= |\tilde{\bf M}_{\eta_3} | $\\
                   &          &  & $ > |\tilde{\bf M}_{\eta_4} | = |\tilde{\bf M}_{\eta_5} |= |\tilde{\bf M}_{\eta_6} |$
                   \\
\end{tabular}
\end{ruledtabular}
\end{table}

Figures~\ref{fig13}(c) and \ref{fig13}(d) show the results for ${\bf B}\parallel [111]$ ($B=0.6$), where the system undergoes the successive transitions as $1Q\to3Q\to1Q\to$ paramagnet while increasing temperature. 
In this case, the nonzero $|\tilde{\bf M}_\eta|$ in the $1Q$ phase is chosen from $\eta=1$, $2$, or $3$, 
while those in the $3Q$ phase are limited to the combination of $\eta=4$, $5$, and $6$. 
Note that $\eta$ in $ \{1,2,3\}$ are equivalent under ${\bf B}\parallel [111]$, and the same holds for 
$\eta $ in $\{4,5,6\}$ (see Fig.~\ref{fig08}).
The order parameters in each phase are summarized in Table~\ref{tab2}.
In this case also, $C$ shows delta-function type anomalies and a jump associated with the discontinuous and continuous transitions, respectively, 
and $\chi$ becomes nonzero in the $3Q$ phase taking a much larger absolute value than that in the $3Q$ model, as shown in Fig.~\ref{fig13}(d).
The large $|\chi|$ is again due to the topological nature of the stacked SkXs with $N_{\rm sk}=-1$.

Next, let us discuss the results for $(D,\Delta)=(0.2,0.3)$ in Fig.~\ref{fig12}. 
In this case, at zero magnetic field, the $6Q$ state persists up to the transition to the paramagnet at $T\simeq 1.10$. 
When we apply the magnetic field, it remains stable in the low field region, 
but turns into the $3Q$ phase in the entire temperature range regardless of the magnetic field direction. 
With further increasing the magnetic field, 
however, the system behaves differently: 
While there is no other ordered phase for ${\bf B}\parallel [100]$ [Fig.~\ref{fig12}(a)], 
we find an additional first-order phase transition to the $1Q$ phase for ${\bf B}\parallel [110]$ [Fig.~\ref{fig12}(b)], and two additional ones to the $3Q'$ and $6Q$ phases for ${\bf B} \parallel [111]$ [Fig.~\ref{fig12}(c)]. 
The case of ${\bf B} \parallel [111]$ is particularly interesting as it shows reentrant transitions from $6Q$ to $3Q$ and $3Q'$, and to $6Q$ while increasing the magnetic field.

Figure~\ref{fig14} shows the field dependences of $|\tilde{\bf M}_\eta|$, $m$, $\chi$, and $N_{\rm sk}$ at $T=0.2$. 
First, for ${\bf B}\parallel [100]$, while increasing the magnetic field,
the system undergoes a first-order phase transition from the $6Q'$ phase, 
which has a different distribution of $|\tilde{\bf M}_\eta|$ from the $6Q$ phase in Fig.~\ref{fig11}(a) (see Table~\ref{tab2}; see also Sec.~\ref{sec:hidden_transition}), 
to the $3Q$ phase with clear jumps of $|\tilde{\bf M}_\eta|$, as shown in Fig.~\ref{fig14}(a). 
The discontinuity is also found for $m$ and $\chi$, as shown in Fig.~\ref{fig14}(b). 

It is worthy noting that while $\chi$ is nonzero in the $6Q'$ phase, 
the absolute value is much smaller than that in the $3Q$ phase.
The value of $\chi$ in the $3Q$ phase is comparable to that in Figs.~\ref{fig13}(b) and \ref{fig13}(d), 
because this $3Q$ state is also topologically nontrivial with $N_{\rm sk}=-1$, 
as shown in Fig.~\ref{fig14}(b). 
In this case, the solid angle $\Omega$ is calculated on the $(100)$ slice.
In the $3Q$ phase, however, we find a topological transition from $N_{\rm sk}=-1$ to $0$ at $B\simeq 1.104$, 
where $\chi$ is rapidly suppressed, as shown in Fig.~\ref{fig14}(b). 
We show the spin configurations and the distributions of the solid angle on the $(100)$ slice for the $N_{\rm sk}=-1$ and $N_{\rm sk}=0$ states in Figs.~\ref{fig15}(a) and \ref{fig15}(b), respectively. 
The green rhombi correspond to a $(100)$ slice of the magnetic unit cell on which $\Omega$ and $N_{\rm sk}$ are computed; 
note that the spin structure does not change along the $[111]$ direction in this $3Q$ state. 
The change of $N_{\rm sk}$ is mainly caused by 
changes of the spin configurations near the triangular plaquettes having large $|\Omega|$ [three blue triangles in Fig.~\ref{fig15}(b)]:
These plaquettes exhibit sign change of $\Omega$ before and after the topological transition.

With a further increase of the magnetic field, the system continuously changes into the paramagnet at $B\simeq 3.031$, as shown in Figs.~\ref{fig14}(a) and \ref{fig14}(b). 
Similar to the previous case from the $6Q$ phase to the paramagnet in Figs.~\ref{fig13}(a) and \ref{fig13}(b),
the three components of the order parameters $|\tilde{\bf M}_\eta|$ show different critical behaviors: 
$|\tilde{\bf M}_{\eta_1}| =|\tilde{\bf M}_{\eta_2}| \propto |B-B_c|^{1/2}$ and $|\tilde{\bf M}_{\eta_3}| \propto |B-B_c|$.
This behavior is also understood from the Ginzburg--Landau type argument:
In this case, the third order terms like $B [ \tilde{M}^x_{\eta_3} (\tilde{\bf M}_{\eta_1} \cdot \tilde{\bf M}_{\eta_2})^*+{\rm c.c.} ]$,
mainly contributes to stabilize the $3Q$ phase. 
Here, $\{\eta_1, \eta_2, \eta_3\}$ are chosen to satisfy ${\bf Q}_{\eta_1} \pm {\bf Q}_{\eta_2} \pm {\bf Q}_{\eta_3} =0$, 
and in addition,
$\eta_1$ and $\eta_2$ are chosen from $1$, $3$, $4$, or $6$, for which the easy axis in the corresponding symmetric interaction is perpendicular to ${\bf B}$, 
and $\eta_3$ is chosen from $2$ or $5$, for which the easy axis is parallel to ${\bf B}$ (see Fig.~\ref{fig08} and Table~\ref{tab2}). 
Thus, in this transition, $\tilde{\bf M}_{\eta_1}$ and $\tilde{\bf M}_{\eta_2}$ are the primary order parameters, acting as internal fields to induce the secondary one $\tilde{\bf M}_{\eta_3}$.

Next, for ${\bf B}\parallel [110]$, the system exhibits successive phase transitions as $6Q\to 3Q\to 1Q\to$ paramagnet, as shown in Fig.~\ref{fig14}(c). 
The transitions between the $6Q$ and $3Q$ phases and between the $3Q$ and $1Q$ phases are both of first order, the former of which is similar to that for ${\bf B}\parallel [100]$, 
while the transition from the $1Q$ phase to the paramagnet is of second order. 
As shown in Fig.~\ref{fig14}(d), $\chi$ is nonzero in both $6Q$ and $3Q$ phases, 
but the absolute value is much larger in the $3Q$ phase due to the nonzero $N_{\rm sk}$ as for the case of ${\bf B}\parallel [100]$. 
In this case, however, $N_{\rm sk}$ is $-1$ in the entire region of the $3Q$ phase, and there is no topological transition in the $3Q$ phase, in contrast to the ${\bf B}\parallel [100]$ case in Figs.~\ref{fig14}(a) and \ref{fig14}(b). 
We note that there is a sign change in $\chi$ in the $6Q$ phase, which will be discussed in Sec.~\ref{sec:hidden_transition}.

Finally, for ${\bf B}\parallel [111]$, the system exhibits reentrant transitions as $6Q\to 3Q\to 3Q'\to 6Q\to$ paramagnet, as shown in Fig.~\ref{fig14}(e).
The $3Q'$ phase is distinguishable from the $3Q$ phase as these phases have different combinations of the nonzero $\tilde{\bf M}_\eta$:
$|\tilde{\bf M}_4| = |\tilde{\bf M}_5| = |\tilde{\bf M}_6| >0$ in the $3Q$ phase, 
while $|\tilde{\bf M}_{\eta_1}| = |\tilde{\bf M}_{\eta_2}| \geq |\tilde{\bf M}_{\eta_3}| > 0$ with $\{\eta_1,\eta_2,\eta_3\} = \{1,2,6\}$, $\{1,3,5\}$, and $\{2,3,4\}$ in the $3Q'$ phase; 
see Table~\ref{tab2}.
Similar to the previous two cases of ${\bf B}\parallel [100]$ and ${\bf B}\parallel [110]$, all the phase transitions are of first order, 
except for that to the paramagnet. 
Moreover, similar to the case of ${\bf B}\parallel [100]$,
the system exhibits a topological transition within the $3Q$ phase at $B\simeq 2.093$, 
where $N_{\rm sk}$ changes from $-1$ to $0$ and $\chi$ is rapidly suppressed, as shown in Fig.~\ref{fig14}(f). 
The spin configurations and the distributions of the solid angle for the $N_{\rm sk}=-1$ and $0$ states are shown in Figs.~\ref{fig16}(a) and \ref{fig16}(b), respectively.
Similar to the case of ${\bf B}\parallel [100]$, the plaquettes with large $|\Omega|$ exhibit sign changes of $\Omega$, 
which mainly contributes to the change of $N_{\rm sk}$. 
In this case also, we note that there is a sign change in $\chi$ followed by the small but nonzero $N_{\rm sk}$ in the low-field $6Q$ state. 
We will touch on this issue in Sec.~\ref{sec:hidden_transition}.


\subsection{Remarks on hidden transitions}\label{sec:hidden_transition}
In this section, we describe two possible types of hidden phase transitions that were found through the present analysis.
One is associated with phase shifts in the complex variables $\tilde{M}^\alpha_\eta$, namely, changes in $\arg \tilde{M}^\alpha_\eta$, 
and the other is associated with changes in the distribution of the amplitudes $|\tilde{\bf M}_\eta|$ while keeping the set of nonzero $|\tilde{\bf M}_\eta|$. 
We note that the importance of the former type has recently been pointed out in both experiment and theory~\cite{Kurumaji2019,Shimizu2021b,Hayami2021c}.
Both types of the transitions are obscure, showing very weak anomalies in the physical quantities, 
and therefore, it is hard to trace them throughout the phase diagrams. 
Thus, we do not indicate such hidden transitions on the phase diagrams shown above. 

For the former type of the hidden transitions, we find, at least, four examples.
For all the cases, the transitions are of first order.
The first case is in the $3Q$ phase in the ground-state phase diagram of the $3Q$ model~(Fig.~\ref{fig02}).
We find that the phases of all $\tilde{M}^\alpha_\eta$ are odd multiples of $\pi/\Lambda$ for small $D$,
namely, $\arg \tilde{M}^\alpha_\eta = \pi n^\alpha_\eta/\Lambda$ with an odd integer $n^\alpha_\eta$, 
while they are even multiples of $\pi/\Lambda$ for large $D$.
The second case is found in the temperature evolution of the $2Q'$ phase in Fig.~\ref{fig06}, 
indicated by small jumps of $\tilde{\bf M}_1$ and $\tilde{\bf M}_2$ near $T \simeq 0.1$.
In this transition, similar phase shifts occur for the nonzero components of the order parameters,
$\tilde{M}^y_1$ and $\tilde{M}^z_1$: 
Both $n^y_1$ and $n^z_1$ are odd integers in the low $T$ regime, 
while they become even integers in the high $T$ regime.
The third and fourth cases are found in the field evolution of the low-field $6Q$ phase of the $6Q$ model
under ${\bf B}\parallel [110]$~[Figs.~\ref{fig14}(c) and \ref{fig14}(d)] and ${\bf B}\parallel [111]$~[Figs.~\ref{fig14}(e) and \ref{fig14}(f)].
In both cases, $\tilde{M}^\alpha_\eta$ exhibits a small jump, 
and $\chi$ changes its sign discontinuously.
In the case of ${\bf B}\parallel [111]$, this transition is followed by a small change in $N_{\rm sk}$, 
as shown in Fig.~\ref{fig14}(f), which is caused by motions of hedgehogs and antihedgehogs. 
The onset of $N_{\rm sk}$ appears at a slightly higher $B$ than the discontinuous transition.
In contrast to the former two cases, 
we cannot determine precisely the phase shifts in these cases because the phase shifts occur in the subdominant components of $\tilde{\bf M}_\eta$ and are difficult to follow within the present numerical accuracy. 

We note that the phase transitions associated with phase shifts can take place in the system with $N_Q$ is larger than the spatial dimension $d$ in the continuum limit, 
which corresponds to the large $\Lambda$ limit in the present models on the discrete lattice. 
This is because a phase shift is reduced to a spatial translation when $N_Q \leq d$ in the continuum limit~\cite{Shimizu2021b}.
Our $3Q$ model is marginal as $N_Q=d=3$, 
and hence, 
we expect that the hidden transitions in the former two cases above may disappear when $\Lambda$ is increased. 
Meanwhile, our $6Q$ model satisfies the condition as $N_Q=6 > d=3$, and therefore, the latter two transitions may survive even in the large $\Lambda$ limit.

For the latter type of the hidden transitions, we find only one example in the current analysis.
It takes place in the low-field $6Q$ phase for ${\bf B}\parallel [100]$ in Figs.~\ref{fig14}(a) and \ref{fig14}(b).
In this transition, the distribution of $|\tilde{\bf M}_\eta|$ appears to change from $|\tilde{\bf M}_{\eta_1}| > |\tilde{\bf M}_{\eta_2}| =|\tilde{\bf M}_{\eta_3}| = |\tilde{\bf M}_{\eta_4}|=|\tilde{\bf M}_{\eta_5}|>|\tilde{\bf M}_{\eta_6}|$ 
to $|\tilde{\bf M}_{\eta_1}| > |\tilde{\bf M}_{\eta_2}| =|\tilde{\bf M}_{\eta_3}| > |\tilde{\bf M}_{\eta_4}|=|\tilde{\bf M}_{\eta_5}|>|\tilde{\bf M}_{\eta_6}|$ at $B\simeq 0.21$, 
where $\{\eta_1, \eta_2, \eta_3,\eta_4,\eta_5,\eta_6\} = \{2, 4,6,1,3, 5\}$ or $\{5,1,3,4,6,2\}$. 
The distribution changes continuously, and $\chi$ shows a peak near the change.
Thus, this transition looks continuous, while the possibility of crossover or weak first-order phase transition cannot be ruled out due to less accuracy.

\section{Summary and perspectives}\label{sec:summary}
To summarize, we have developed a theoretical framework to investigate the phase competition between multiple-$Q$ magnetic orders in a class of effective spin models with long-range magnetic interactions derived from the coupling to conduction electrons. 
In addition, applying the framework to two models, the $3Q$ and $6Q$ models, we have elucidated the magnetic field-temperature phase diagrams, 
which reveal a variety of interesting magnetic and topological transitions.

Specifically, 
we constructed two methods, method I and method II, 
both of which are based on the steepest descent method and provide the exact solutions in the thermodynamic limit. 
They are complementary to each other:
Method I is computationally cheap but limited to two-spin interactions, while method II is computationally expensive but can be applied to more generic multiple-spin interactions. 
The framework is unbiased and concise, and has many advantages over previously used methods, such as variational calculations and Monte Carlo simulations.

Using the framework, we studied the ground-state and finite-temperature phase diagrams of the $3Q$ and $6Q$ models on a simple cubic lattice in an external magnetic field applied to the $[100]$, $[110]$, and $[111]$ directions. 
The models include the anisotropic symmetric interactions and the DM-type antisymmetric interactions, 
and exhibit multiple-$Q$ magnetic orderings in the ground states. 
By detailed analysis of the ground-state spin configurations at zero magnetic field, 
we found magnetic hedgehogs and antihedgehogs forming 3D lattices in the $3Q$ phase of the $3Q$ model and the $6Q$ phase of the $6Q$ model;
we also found magnetic skyrmions forming 2D lattices in the $3Q$ phase of the $6Q$ model.
By further analysis with introducing temperature and the external magnetic field, we obtained the complete phase diagrams with a higher resolution than ever before.

We found two particularly interesting features in the phase diagrams:
Thermally-stabilized multiple-$Q$ spin states and topological transitions in the multiple-$Q$ phases.
As the former features, we found that a $3Q$ phase of the $3Q$ model and a $6Q$ phase of the $6Q$ model appear only at finite temperature~[Figs.~\ref{fig04}(c) and \ref{fig11}(a)].
The detailed analysis by the Ginzburg--Landau expansion indicates that the magnetic field plays also an important role for the stabilization of the $6Q$ phase,
while the $3Q$ phase is stable even at zero magnetic field.
As the latter features, we found a transition between the $3Q$ state composed of stacked skyrmion crystals and the $6Q$ state with a hedgehog lattice in the $6Q$ model~[Figs.~\ref{fig09}, \ref{fig11}(a), \ref{fig12}, \ref{fig13}(a), \ref{fig13}(b), and \ref{fig14}]. 
We also found topological transitions within each $3Q$ and $6Q$ phase, 
where the skyrmion number vanishes while changing temperature and the magnetic field~[Figs.~\ref{fig13}(b), \ref{fig14}(b), and \ref{fig14}(f)].

Our results demonstrate that the newly developed framework in this paper provides a powerful tool to investigate the phase competition in the effective spin models for magnetic metals. 
It can be applied to a generic form of the Hamiltonian, which includes not only two-spin but also multiple-spin interactions with any anisotropy. 
In recent years, several variations of such effective spin models have been studied for understanding of multiple-$Q$ magnetic orderings in many materials, e.g., the skyrmion lattices in 
Gd$_2$PdSi$_3$~\cite{Yambe2021}, 
GdRu$_2$Si$_2$~\cite{Yasui2020,Hayami2021d,Khanh2022}, 
Gd$_3$Ru$_4$Al$_{12}$~\cite{Hirschberger2021},
and EuPtSi~\cite{Hayami2021g},
and the hedgehog lattices in 
MnSi$_{1-x}$Ge$_x$~\cite{Okumura2020,Shimizu2021a,Kato2021}.
Our framework would be useful to clarify the complete phase diagrams and the nature of the transitions between different multiple-$Q$ phases in a high resolution. 
While our demonstration was limited to the model with two-spin interactions only, the models with higher-order spin interactions, such as biquadratic and bicubic interactions, are important for a new generation of the topological multiple-$Q$ magnetic orderings that appear in the systems with no or less anisotropy arising from the spin-orbit coupling~\cite{Hayami2021b}.
Such extensions are left for future studies.

\begin{acknowledgments}
This work was supported by Japan Society for the Promotion of Science (JSPS) KAKENHI Grant Nos.
JP18K03447 and JP19H05825 
and JST CREST Grant No. JPMJCR18T2.
\end{acknowledgments}


\bibliography{draft} 

\end{document}